\documentclass[fleqn,usenatbib]{mnras}

\usepackage{newtxtext,newtxmath}
 
\usepackage[T1]{fontenc}

\DeclareRobustCommand{\VAN}[3]{#2}
\let\VANthebibliography\thebibliography
\def\thebibliography{\DeclareRobustCommand{\VAN}[3]{##3}\VANthebibliography}

\usepackage{graphicx}	
\usepackage{amsmath}	
\usepackage{amssymb}	
\usepackage{float}      
\usepackage[utf8]{inputenc}

\title[Census of LISA MBHB host galaxies in SDSS]{Preliminary census of galaxies in the LISA localisation volume: I. Searching for LISA candidate massive black hole binary merger hosts using Sloan Digital Sky Survey photometry}

\author[Carolyn L. Drake et al.]{Carolyn L. Drake$^{1}$, Jessie Runnoe$^{1, 2}$, Aaron Stemo$^{1,2}$, Tamara Bogdanović$^{3}$, Michael Eracleous$^{4}$, John Ruan$^{5}$
\newauthor Kaitlyn Szekerczes$^{4}$, Kate Futrowsky$^{3}$, Weixiang Yu$^{5}$
\\
$^{1}$ Vanderbilt University, Department of Physics \& Astronomy, 6301 Stevenson Center, Nashville, TN 37235, USA
\\
$^{2}$ Fisk University, Department of Life and Physical Sciences, 1000 17th Avenue N, Nashville, TN 37208, USA
\\
$^{3}$ Center for Relativistic Astrophysics, School of Physics, Georgia Institute of Technology, Atlanta, GA 30332
\\
$^{4}$ Department of Astronomy \& Astrophysics, and Institute for Gravitation and the Cosmos, Penn State
\\
University, 525 Davey Lab, 251 Pollock Road, University Park, PA 16802, USA
\\
$^{5}$ Department of Physics \& Astronomy, Bishop's University, Sherbrooke, QC J1M 1Z7, Canada}

\date{Accepted XXX. Received YYY; in original form ZZZ}

\pubyear{2025}

\begin{document}
\label{firstpage}
\pagerange{\pageref{firstpage}--\pageref{lastpage}}
\maketitle

\begin{abstract}
With the launch of the Laser Interferometer Space Antenna (LISA), we will be able to estimate the sky position, luminosity distance (d$_{L}$), chirp mass, and mass ratio for detected merging massive black hole binary (MBHB) systems. LISA’s uncertainties on these estimates will evolve over time, and enable electromagnetic (EM) follow-up observations as early as a month from coalescence. In this paper, we create a framework that takes simulated LISA parameter estimates for sky localisation and d$_{L}$ for a MBHB and performs a census of matching EM galaxies, or candidate host galaxies. We used this framework to investigate these parameter estimates for simulated MBHB systems with masses of $3\times10^{5}$, $3\times10^{6}$, and $1\times10^{7}$~M$_{\odot}$ at redshifts of $0.3$ and $0.5$ and used these parameters to select matching galaxies from archival Sloan Digital Sky Survey (SDSS) photometry. We found that the number of candidate host galaxies for a simulated MBHB system at a redshift of $0.3$ and $1$ hour from coalescence ranged from tens to thousands. After coalescence, we found that our census numbers dropped to zero for all systems when considering median constraints most likely due to survey limitations. For a MBHB with mass $3\times10^{6}$~M$_{\odot}$ at $1$ hour from coalescence, increasing the redshift from $0.3$ to $0.5$ or varying the sky position within the SDSS footprint resulted in the number of EM counterparts increasing by approximately a factor of $2$.
\end{abstract}




\section{Introduction}
\label{sec: intro}
The Laser Interferometer Space Antenna \citep[LISA;][]{amaroseoane2017laser, colpi2024lisadefinitionstudyreport} is a space-based gravitational wave (GW) observatory expected to launch in the $2030$s. LISA will be sensitive to GW frequencies of $0.1$~mHz$-1$~Hz and will be capable of detecting both local and extragalactic GW sources. The loudest sources in the LISA band are expected to be merging massive black hole binaries (MBHB) with a total mass in the range of $10^{5}-10^{7}$~M$_{\odot}$ \citep{Amaro_Seoane_2023}. These merging MBHBs will also be detectable by LISA out to high redshifts \citep[$\lesssim20$; ][]{colpi2019}. MBHBs are thought to form as a result of galaxy mergers \citep{begelmanetal_1980}, but LISA will capture them immediately prior to, and at, coalescence \citep{kleinetal2016}. Thus, these MBHB merger events are expected to be prime multi-messenger targets for coordinating LISA GW detections with coincident electromagnetic (EM) or particle observations.


The Laser Interferometer Gravitational-Wave Observatory \citep[LIGO;][]{LIGO} has demonstrated the power of multi-messenger detections by connecting short gamma-ray bursts, previously of unknown origin, to binary neutron star mergers \citep{BNS, Abbott_2017, savchenko17, goldstein17}. The first GW detections of MBHBs by LISA will be groundbreaking, but coincident EM detections can provide crucial additional insights. Such multi-messenger detections could serve as a novel probe of cosmology through independent measurements of redshift and luminosity distance (d$_{L}$) \citep{Arun_2009, Tamanini_2017}. Furthermore, joint detections will enable studies of the dynamics of MBHBs \citep[e.g.,][]{bonettietal_2019} and the role of circumbinary gas in their evolution \citep[e.g.,][]{Escala_2005_gas_does_work, dotti_2007_gas_does_work, lodato_2009_gas_does_work_intricate, chapon_2013_maybe_gas_doesnt_always_work, role_of_gas_in_evo}.


The most likely method by which the first multi-messenger observations of MBHBs will be achieved is by connecting their LISA GW signal with an associated EM counterpart. For every MBHB that LISA detects, analysis of the GW signal will constrain system parameters, including sky position, as a function of time from coalescence, enabling rapid follow-up EM searches for their associated counterparts prior to coalescence. Specifically, LISA's tumbling orbit \citep{Cutler_1998_oldlisadesign, langetal2008} will enable the constraint on the sky position of the GW source to be $\lesssim10$~deg$^{2}$ as early as one month from coalescence \citep{mangiagli20}. Furthermore, the detected GW signal will also enable the estimation of the source's mass ratio, chirp mass, and luminosity distance \citep{hughes2002, bertietal2005, mangiagli20} through the analysis of the gravitational waveforms. These parameters become more constrained as the merger progresses, although the allowed range of relative uncertainties increases as the system nears coalescence. Still, as the MBHB's orbit decays, pinpointing the EM counterpart on the sky becomes a more feasible task.

There are a number of possible EM signatures that may indicate the presence of a MBHB. Throughout the merger process, if one or both black holes are accreting from a surrounding gas reservoir, the system may be identified as an active galactic nucleus (AGN) in its host dwarf galaxy \citep{reinesetal_2013_agnindwarfs, Baldassare_2017_agnindwarfs, Molina_2021_agnindwarfs}. In addition to persistent accretion, transient events may occur at the time of coalescence, or shortly thereafter. For instance, a change in the Poynting flux vector may drive an observable transient flare in the EM, possibly due to Doppler boosting effects associated with coalescence or through an induced heavy inflow of gas on to the binary \citep{Palenzuela_2010, Giacomazzo_2012, Moesta_2012, Kelly_2017}. However, there may be considerable contamination from other EM transients \citep{Yu_2025}. At coalescence of the MBHB system, if the GW emission is asymmetric it can impart a velocity `kick' to the remnant black hole, which will travel through the surrounding gaseous medium and can emit an observable EM afterglow \citep{Lippai_2008, Schnittman_2008, Corrales_2010, Rossi_2010, Meliani_2017}. 

However, these EM signatures depend on the presence of gas and accretion, neither of which is guaranteed. In simulations, the resolution and initial conditions greatly influence whether a simulated MBHB system will form a gas reservoir that accretes onto one or both black holes, forming an AGN \citep{limaetal_2020}. Complementary observations further find that it is not clear that mergers of galaxies guarantee that gas will make its way to the central engine to enable accretion, although it may be more common that it does \citep{ellison_2011, Ellison_2013}. Even if the MBHB is accreting prior to merger, the accretion rate may be extremely low \citep{bodeetal2010, Bogdanovic_2011}. A majority of the black hole population in the local universe is accreting with a radiative efficiency of less than $0.1$ \citep{ho_etal_2009}, meaning that is likely an accreting LISA MBHB will be fairly faint. 

Possible AGN obscuration is another cause for concern. At the luminosities we expect for a LISA MBHB AGN \citep{lopsetal2023}, there is a particularly high incidence of AGN obscuration in the optical band due to dust \citep{merlonietal2013, hickoxetal2018}. Thus, prior identification of active LISA MBHBs via AGN optical luminosity, even if they are accreting at an appreciable rate, can be extremely difficult. 



 

In light of these considerations, it is possible that the EM counterpart to a LISA MBHB merger will instead be the host galaxy. LISA MBHB systems are likely to predominantly reside in smaller galaxies in the stellar mass range of $10^{8}-10^{10}$~M$_{\odot}$ \citep{izetal2023}, corresponding to the dwarf galaxy population \citep{greeneetal_2020_dwarfgalmassrange,Reines_2022_dwarfgalmassrange}. These host galaxies are faint (as faint as absolute magnitude $M_{r, AB}\sim-19$ for the given dwarf mass range in the NASA-Sloan Atlas, see \citealt{blanton_2011_nsatlas}) but detectable in the local universe with large-area sky surveys \citep{lowmassgals_sdss_bradfordetal_2015, lowmassglas_sdss_sunetal_2020,lowmassgals_WISE_jarrettetal_2023}. Treating the host galaxy as the EM counterpart of MBHBs is agnostic to accretion rates and obscuration. This approach complements efforts to find the EM counterpart among the transient source events described above.

The number of AGN or galaxies in the localisation region for a LISA-detected MBHB is expected to be large. Several groups have performed a census exercise by counting the EM sources of different classifications that are consistent with the parameters estimated from the LISA GW signal of a merging MBHB. This exercise was done using a combination of simulation and statistical techniques. Recent work found that in cosmological simulations, close to coalescence, there are $10$s$ - 100$s of possible low-redshift $(z<0.5)$ galaxies that could host the MBHBs LISA detects \citep{lopsetal2023, izetal2023}. Other approaches using statistical techniques, such as luminosity density functions and average galaxy densities, have found anywhere from $10$s to $1000$s of galaxies per square arcminute which agrees with the estimated LISA parameters, at a range of redshifts \citep{Holz_2005, Kocsis_2008}. The variance in the results of these previous census exercises is, in part, due to the evolving LISA design plan \citep{Cutler_1998_oldlisadesign, colpi2024lisadefinitionstudyreport} along with increasingly robust LISA parameter estimation techniques \citep{Vecchio_2004_oldparamest, mangiagli20}. Still, both predictive methods estimate that there will likely be a large number of EM sources that are consistent with a LISA target. Further, these methods do not directly translate to actual EM follow-up strategies as they do not use real individual targets in their approach. Therefore, methods to reduce the number of possible counterparts need to be developed using real astronomical data.

In this paper, we focus on performing a census of extended objects that agree with LISA parameters based on EM surveys. Specifically, we combine various simulated LISA parameter estimations with Sloan Digital Sky Survey (SDSS) data to perform a census of extended objects that are similar to what would be expected of a host galaxy for a LISA MBHB system. Throughout this paper, we refer to these extended objects as candidate host galaxies. This style of census acts as a more realistic proxy of future EM searches for counterparts to LISA sources in comparison to previous efforts that used a simulation-based or analytical approach.


This paper is organized as follows. In Section~\ref{sec: section 2}, we address how we gathered our archival survey data and created a framework that applies LISA parameter estimations to SDSS data. The completed censuses of candidate host galaxies are presented in Section~\ref{sec:Results}. A discussion of our results and their implications is presented in Section~\ref{section: discussion}. Throughout this paper we assume a cosmology consistent with Planck18 \citep{Planck18} with parameters of $H_{0}=(67.66\pm0.42)$~km~s$^{-1}$~Mpc~$^{-1}$, $\Omega_{\Lambda}=(0.69\pm0.01)$, and $\Omega_{m, 0}=(0.31\pm0.01$).  


\section{Developing a Modular Framework for Counting Counterpart Galaxy Candidates}
\label{sec: section 2}

The main objective of this work is to quantify the number of observed galaxies that agree with a variety of simulated LISA MBHB systems. We developed a modular framework to take input LISA parameter estimates and apply them to observations of real galaxies in large-area sky surveys. Thus, in this work, we conducted a census of galaxies that may be EM counterparts of the simulated binary system, or candidate host galaxies. This section discusses the creation of the aforementioned framework and our curation of data from an observational survey.

We used the prescription of \citet{mangiagli20} to obtain parameter estimates for LISA. \citet{mangiagli20} simulated parameter estimation curves for a variety of binary systems using a combination of Fisher Matrix and Markov chain Monte Carlo evaluation methodologies based on the LISA sensitivity curve and a variety of gravitational waveform prescriptions. They then created polynomial fits, which we refer to as `analytical fits', for \textit{on-the-fly} parameter estimation as a simulated LISA MBHB system evolves as a function of time. Given a total MBHB system mass, redshift, location in the sky, and time from coalescence, these fits return the median, $68^{th}$, and $95^{th}$ percentile confidence intervals for sky localisation, luminosity distance, chirp mass, and mass ratio. 

\subsection{Archival Electromagnetic Survey Observations}

The first level of the framework compares the sky position, luminosity distance, and total binary mass constraints from the LISA parameter estimations with corresponding EM properties gathered from archival survey data. For the survey data, we adopted data from the Sloan Digital Sky Survey \citep[SDSS;][]{SDSS} which has a number of advantages. SDSS sky coverage is among the largest for surveys of the northern hemisphere, capturing $14,555$ unique square degrees in imaging with a detection limit of m$_{r, AB}=23.1$ \citep{SDSS}. The SDSS footprint, which has a relatively uniform photometric depth, is larger than all LISA localisation regions investigated in this work. A survey of this size allows our framework to probe the distribution of candidate host galaxies at different sky positions. Furthermore, this large coverage area enables future work to directly compare the results of our census with another survey's overlapping data. Finally, many aspects of other modern surveys, including filter properties and data access, are based on the experience gained with SDSS \citep{Ivezic_2019_LSST}. Therefore, basing our framework on SDSS makes it more easily adaptable to other modern surveys. 

The primary data for this work were pulled from the SDSS Data Release 16 \citep{SDSS_dr16} which can be accessed through the SDSS Structured Query Language (SQL) client\footnote{\url{https://skyserver.sdss.org/dr16/en/tools/search/sql.aspx}}. Although this client is web-based, we accessed it asynchronously using the {\tt SciServer} python package. We utilized only photometric data products from SDSS, with the goal of developing methodologies that will be most widely applicable in the era of photometric surveys like the Vera C. Rubin Observatory Legacy Survey of Space and Time \citep[LSST;][]{Ivezic_2019_LSST}.

We used data from the {\tt GalaxyTag} view of SDSS. {\tt GalaxyTag} reports the most popular quantities from the main SDSS photometry table for all objects which have been identified as spatially extended by SDSS. We joined the {\tt GalaxyTag} view with the views that contain photometric redshift estimates ({\tt PhotoZ}) and data from the Wide-Field Infrared Sky Explorer survey \citep[WISE;][]{WISE} ({\tt wise\_allsky}). Note that this process discards objects that do not have all necessary measurements. To ensure the quality of data from SDSS, we performed cuts based on the {\tt clean}, {\tt mode}, and {\tt type} flags. Specifically, we required {\tt clean}~$=1$, indicating that the photometric measurements were not contaminated in any way, {\tt mode}~$=1$, indicating that the reported measurements are from the primary observation of the target (i.e. the best available if there is more than one), and {\tt type} $=3$, indicating that the object has been identified by SDSS as extended. These selection criteria curated a high-quality photometric dataset of spatially extended galaxies with photometric redshifts and WISE detections which formed the primary database for the queries in this work.

\begin{figure*}
    \centering
    \includegraphics[width=\textwidth]{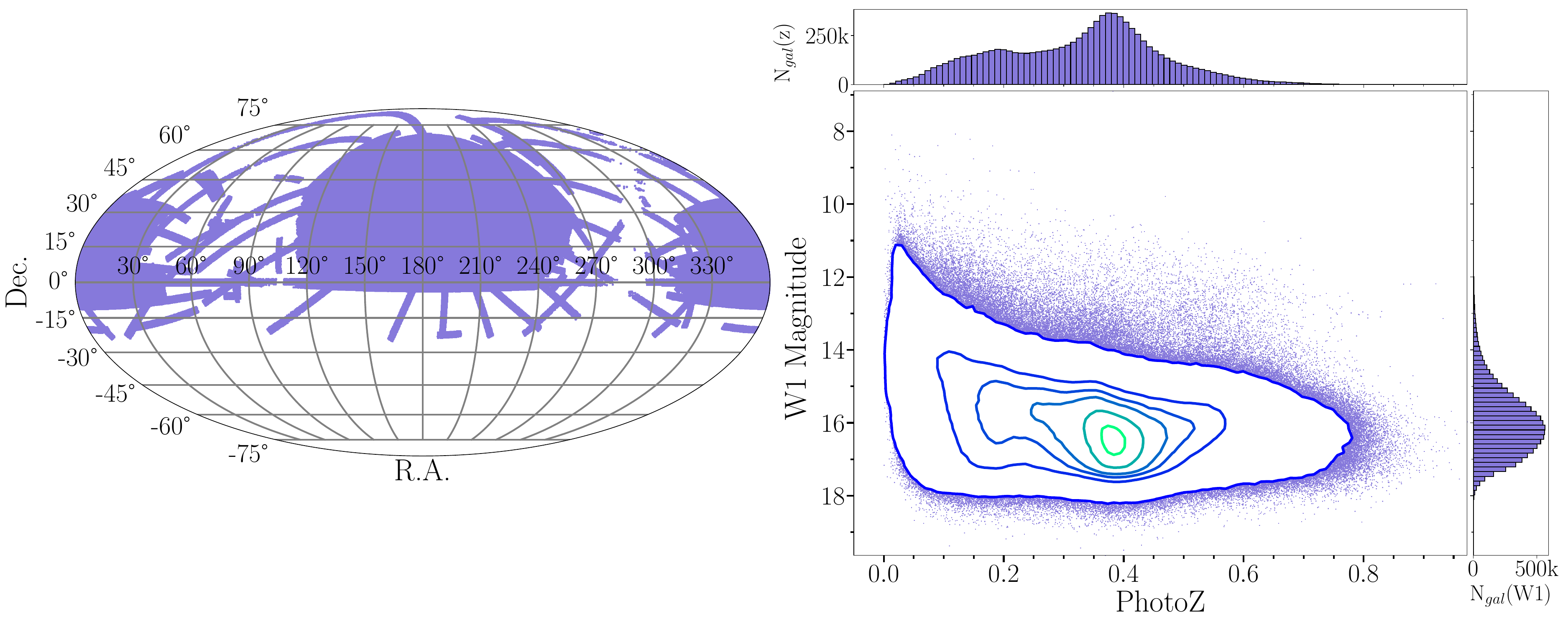}
    \caption{The framework uses data curated from SDSS and WISE of spatially-extended galaxies over a large area of the sky with high quality photometry. Left: A Mollweide projection of the sky distribution of galaxies in the database that agree with our quality cuts; note the large fraction of sky covered by this sample. Right: The $W1$ magnitudes of the galaxies in the curated dataset versus the photometric redshifts from the curated dataset. Distributions for both quantities are appended to their respective axes. The contours encircle roughly $5\%$, $20\%$, $40\%$, $60\%$, $80\%$, and $99.5\%$ of our dataset. Overall, our dataset captures the entirety of the SDSS footprint and peaks at a redshift of $0.4$ and a $W1$ magnitude of $16$.}
    \label{fig:datasets_g}
\end{figure*}

Throughout our framework, we use redshifts for multiple calculations. SDSS has a catalog of photometric redshift measurements contained in the {\tt PhotoZ} view \citep{becketal2016}. These photometric redshift measurements are estimated by a pipeline using a KD-tree nearest neighbor fit process trained on a population of $1,966,000$ spectroscopic galaxies that cover redshifts up to $0.8$. Ultimately, these redshift measurements are used in two ways by our framework. First, these photometric redshift measurements are compared to the LISA luminosity distance constraints to determine if individual galaxies are included in our census. Second, they are incorporated into the mass estimates for the host galaxies. 

To include these photometric redshift measurements in our dataset, we performed the following quality control cuts. We began by removing null values in estimated redshift and redshift error that are reported as $-9999$ in the {\tt PhotoZ} view. We then performed a further cut based on the primary recommendation from \citet{becketal2016} that is predicated on the {\tt PhotoErrorClass} of an object to balance measurement quality with the number of objects which have available measurements. The {\tt PhotoErrorClass} value is determined by each galaxy's colours and photometric errors and represents how well each galaxy's parameters agree with the training set. We adopt redshift estimates that have a {\tt PhotoErrorClass} $=1$ as they are the most accurate and represent over $50\%$ of the sources contained in {\tt GalaxyTag}, {\tt wise\_allsky}, and {\tt PhotoZ}. 

In order to utilize the constrained LISA MBHB system masses, we included a method to estimate mass in our framework that leverages WISE $W1$ magnitudes. WISE is a useful survey as infrared emission is less prone to dust extinction and it traces older stellar populations leading to more robust mass estimates \citep{Jarrett_2013}. WISE has all-sky coverage which makes its implementation broadly useful in combination with any optical survey. We curated the WISE photometry with quality control cuts. We used the profile fitted magnitudes in the $W1$ filter and removed any sources with null values in $W1$ and $W1$ uncertainty. We also required {\tt w1flg}~$=0$ for all objects, indicating no contamination in the measurement reported by WISE.

\begin{figure}
    \centering
    \includegraphics[width=1.05\columnwidth]{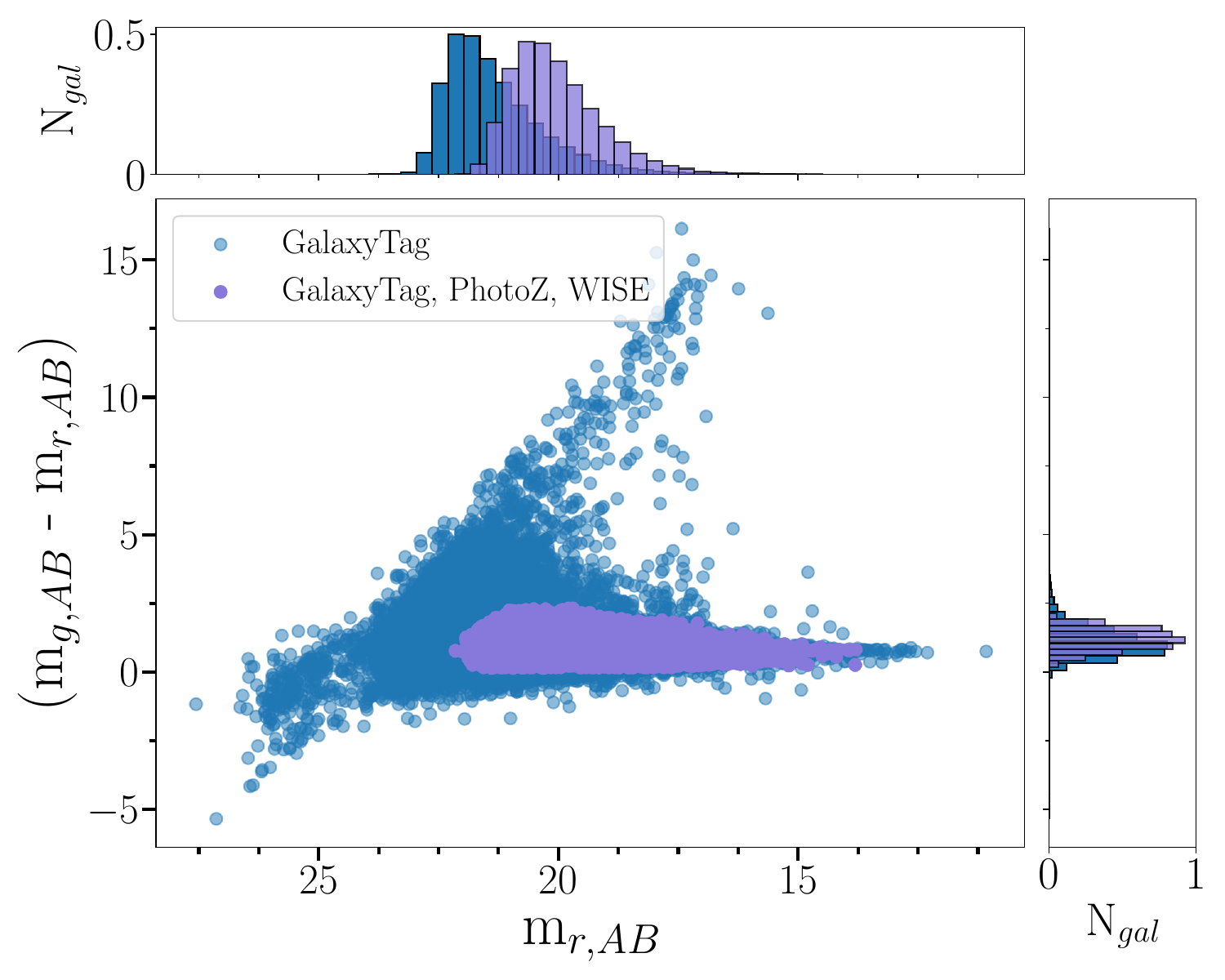}
  \caption{Overall, our curated dataset (purple) matches well with the statistics of the original dataset from {\tt GalaxyTag} (blue), especially in colour space, although we do sample slightly higher luminosities. Top Histogram: Normalized distributions of the apparent magnitude for the curated dataset compared to the raw dataset pulled from {\tt GalaxyTag}. The curated dataset has a slightly lower median magnitude value of $\sim20$ in comparison to the raw dataset which has a median magnitude of $\sim21.5$. Right Histogram: Normalized distributions of the $(m_{g,AB}-m_{r,AB})$ colour for both the curated dataset and the raw dataset included in {\tt GalaxyTag}. Points from our curated dataset generally lie in the center of (and follow the same shape as) the raw dataset. The median colour value of the two distributions only differs by $\sim0.2$. Center Plot: Apparent $m_{r, AB}$ magnitude versus $(m_{g,AB}-m_{r,AB})$ colour for the curated dataset and the raw {\tt GalaxyTag} dataset. Our curated dataset is pulled from the higher luminosity end of $m_{r,AB}$ and from the densest portion of colour space, which follows and agrees with the histograms in the previous plots.}
  \label{fig:xmatch_stats}
\end{figure}

The photometric measurements with all quality control cuts are visualized in Figure~\ref{fig:datasets_g}. The curated dataset consists of over $9,000,000$ sources, which is $\sim14\%$ of the total number of sources without control cuts, and captures the entirety of the SDSS observing footprint. The curated dataset has the highest density of sources at redshift $\sim0.4$ and  $W1$ $\sim16$. We note that joining catalogs with the above quality cuts creates a dataset biased towards a slightly higher luminosity than the raw {\tt GalaxyTag} view. This bias is illustrated in Figure~\ref{fig:xmatch_stats}. However, in all cases, the general shape in both magnitude space and colour of our curated dataset remains consistent with the raw {\tt GalaxyTag} view. Therefore, the cross-matching and quality cuts that we implement seem to not create a significant bias.


\subsection{Sky Position Constraints}

To incorporate sky localisation constraints into the framework, we utilized the function {\tt GetNearbyObjEq} which is provided by the SDSS SQL service. 
{\tt GetNearbyObjEq} creates a circular projection, known colloquially as a sky cap, whose area depends only on its radius. In the framework, the radius of the cap is determined by the LISA MBHB sky localisation constraints that are calculated from the analytical fits. The sky cap then functions as a proxy for the true LISA sky localisation area and was the first step of our census. The sky cap is then centered on a specified point of right ascension (RA) and declination (Dec). Finally, a query returns the $W1$ magnitudes and photometric redshift measurements for all objects that have clean measurements and RA's and Dec's that fall within the cap.

The RA and Dec of each object as reported by SDSS were treated as having a single error value. Positional errors are on the order of arcseconds for SDSS \citep{SDSS}, which translates to an uncertainty in localisation area for RA and Dec of order $10^{-3}$~deg$^{2}$. The uncertainty for both RA and Dec in SDSS are then negligible in comparison to the uncertainties in sky position provided by LISA which are on the order of deg$^{2}$ \citep{mangiagli20}.

This step of our framework was the most general as it was only concerned with whether the object agrees with the sky localisation provided by the LISA parameter estimates. The sky localisation constraint was also carried through to all other steps of our framework, as focusing on objects only included in this sky cap greatly reduced the computation time needed for the rest of the framework. Otherwise, the computation time required by other steps would be intractable.


\subsection{Luminosity Distance Constraints}

\begin{figure*}
    \centering
    \includegraphics[width=0.45\textwidth, trim={0 0 19.05cm 0}, clip]{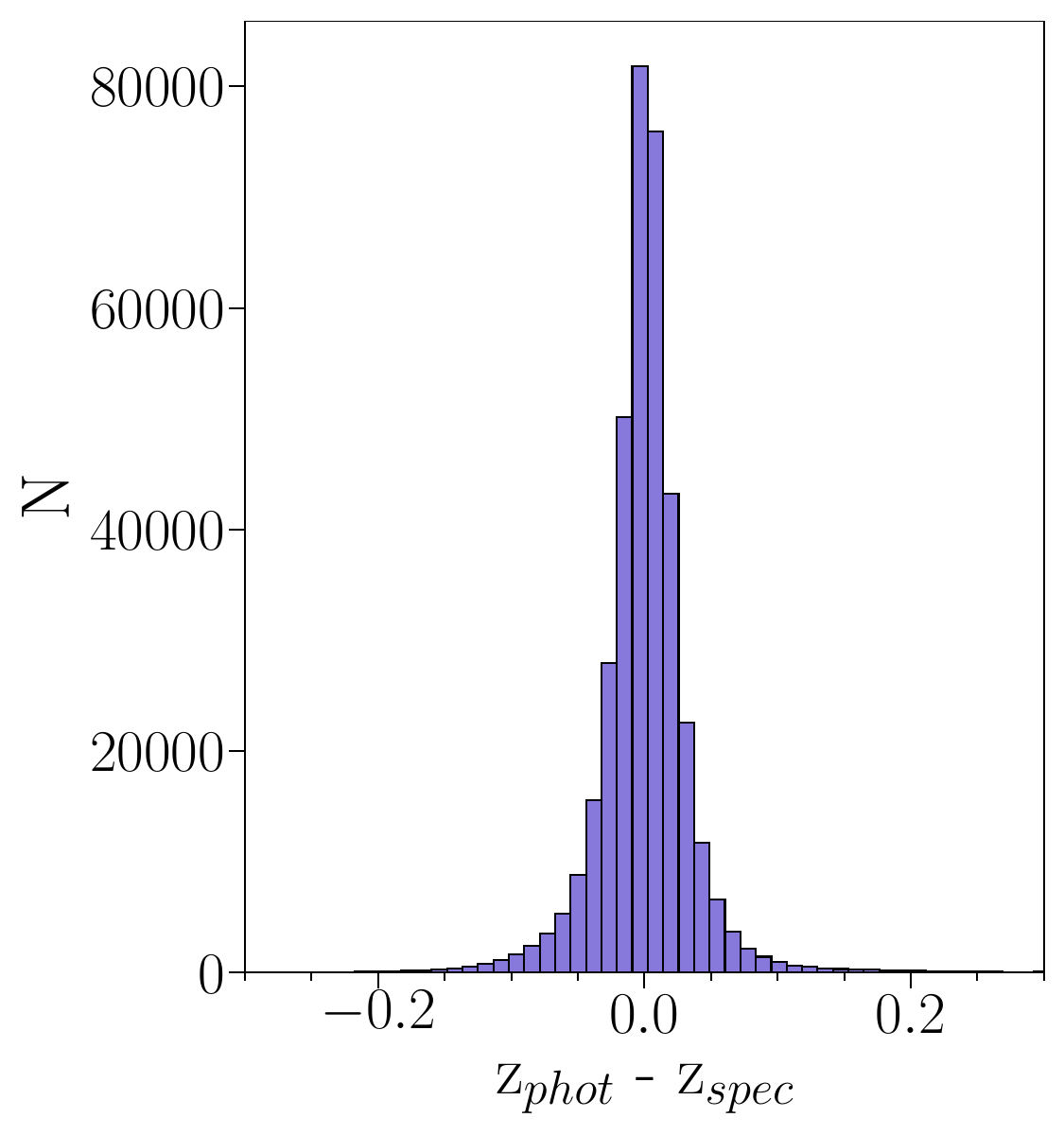}
    \hspace{0.5cm}
    \includegraphics[width=0.45\textwidth, trim={19.05cm 0 0 0}, clip]{hist_photo_v_spec_z_new.pdf}
    \caption{The {\tt GalaxyTag} photometric redshifts provide a good estimate of the galaxy redshift measured from the spectra. There is a cluster of points with overestimated photometric redshifts, but this cluster has little effect on the overall results of this work because it is not in the redshift range investigated in this paper and makes up a negligible portion of the compared data. The distribution of differences has a mean of $\sim0.001$ and a standard deviation of $\sim0.051$ meaning there is relatively low scatter in the difference and that there is not an impactful systematic bias. Left: Distribution of the difference in photometric redshifts and spectroscopic redshifts. Right: Spectroscopic redshifts versus photometric redshifts along with contours encircling $5\%$, $20\%$, $40\%$, $60\%$, and $90\%$ of the sources and the dashed line showing where they are equal. The distribution for each quantity is appended to the appropriate axis.}
    \label{fig:photoz_dataset}
\end{figure*}

To adopt the LISA luminosity distance constraints, we assumed a cosmology and  compared the redshift  of real galaxies to the uncertainties provided by the analytical fits. To assess the degree of uncertainty introduced by adopting photometric redshifts in the framework, we performed a comparison with spectroscopic redshifts for a small sample with both measurements.  The eBOSS FIREFLY value-added catalogue \footnote{\url{https://www.sdss.org/dr18/data_access/value-added-catalogs/?vac_id=47}} \citep{wilkinson_2019_firefly_code, eboss_firefly_VAC} includes objects with {\tt GalaxyTag} photometric measurements and spectroscopic redshifts. Their comparison is shown in Figure~\ref{fig:photoz_dataset}. Overall, we found that the distribution of differences between the photometric and spectroscopic redshifts is centered at zero with a standard deviation of $\sim0.05$.  There is a large population of objects with spectroscopic redshifts near zero, but with overestimated photometric redshifts. This discrepancy has a negligible impact on the results of this work as these overestimated redshifts make up less than $1\%$ of the redshifts that we compared. Thus, we find that there is no impactful systematic bias in the photometric redshifts.

To incorporate the luminosity distance constraints, we treated the uncertainties from \citet{mangiagli20} as strict thresholds for including a galaxy in the census. The analytical fits report percent uncertainties for luminosity distance based on the input redshift of the simulated merging MBHB system. Since SDSS reports photometric redshifts, not luminosity distances, we assumed the cosmology detailed in Section~\ref{sec: intro} and converted the luminosity distance uncertainties to a range of acceptable redshifts.


We compared this range of redshifts to the reported photometric redshift values for all objects that agreed with our quality control cuts and were in the cap determined by the sky localisation constraints. If an object's redshift enters the redshift range found above within $1\sigma$, it was included in our census as a possible candidate host.

\subsection{Mass Constraints}

To introduce a LISA mass constraint into our framework, we compared the chosen total mass of the MBHB that was used in the analytic fits with a photometrically estimated black hole mass hosted by each galaxy in the sky localisation cap. If the photometrically estimated black hole mass agreed within error with the MBHB total mass, it was included in the final census. We detail the procedure below.


We began with the $W1$ photometry values of galaxies that agreed with all listed quality cuts and fell within the localisation cap. We used these  $W1$ magnitudes to calculate the stellar mass of these galaxies. As the sources can have non-negligible redshifts, we applied a k-correction of
\begin{equation}
    k_{W1} = -7.1\times\textrm{log}_{10}(1+z), 
\end{equation}
following \citet{kettlety18} to each $W1$ magnitude when we calculated the flux. We then normalized the $W1$ luminosity for each object to the Solar luminosity in $W1$:
\begin{equation}
    L_{W1} = \frac{4\pi d_{L}^{2}F_{W1}}{L_{W1\odot}}L_{\odot},
\end{equation}
where d$_{L}$ is the luminosity distance based on the reported median photometric redshift for each object, $F_{W1}$ is the flux of the object as found from a k-corrected $W1$ magnitude, and $L_{W1\odot}$ is the Solar luminosity in $W1$, or $1.6 \times 10^{32}$~erg~s$^{-1}$. The total stellar mass of each galaxy is calculated from its $W1$ luminosity by using a $W1$ mass-to-luminosity relationship of $0.2$, which is within the range presented by \citet{leroyetal2019}.  



\begin{figure*}
    \centering
    \includegraphics[width=0.4375\textwidth, trim={0 0 18.5cm 0}, clip]{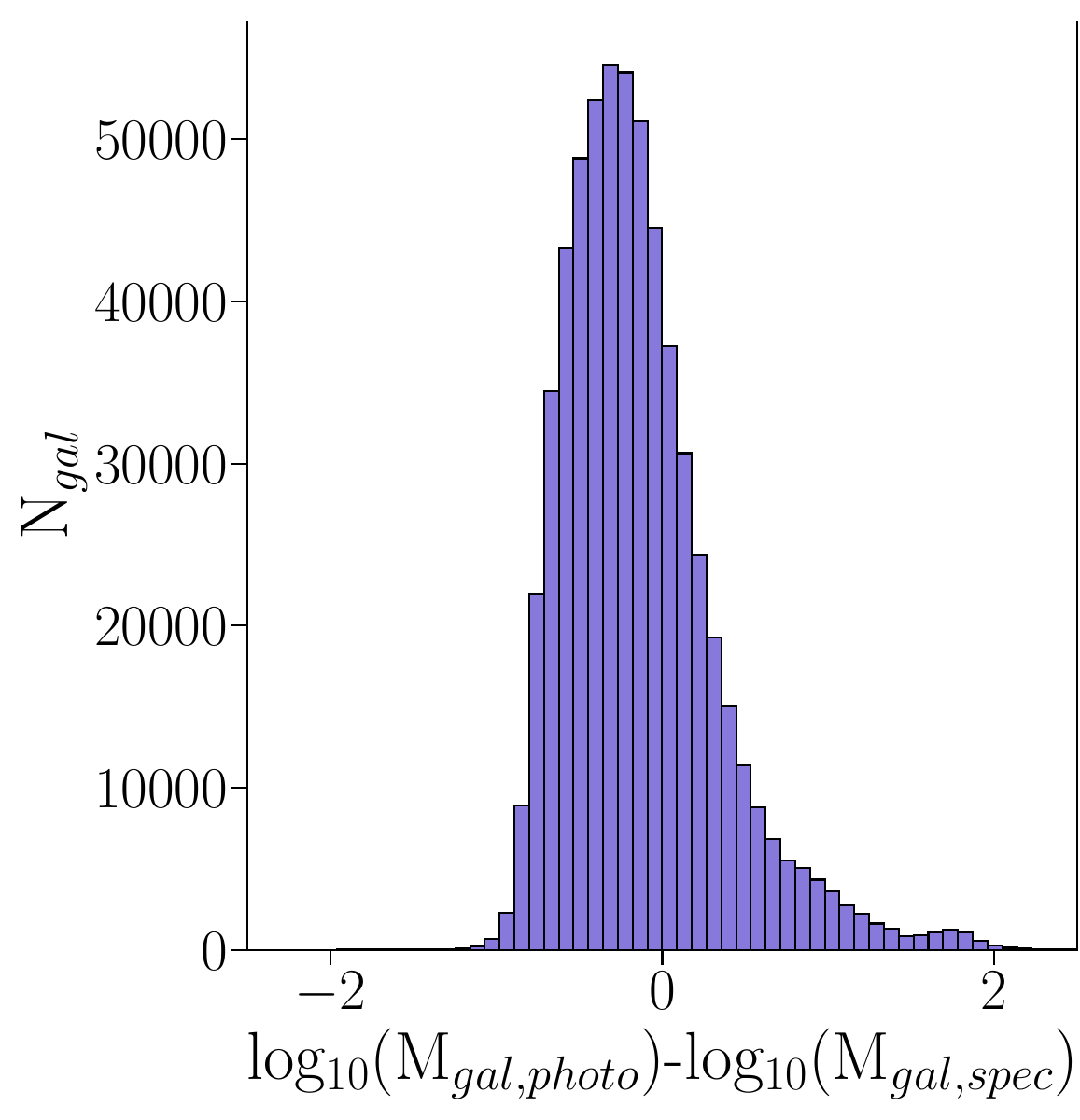}
    \hspace{0.5cm}
    \includegraphics[width=0.41\textwidth, trim={19.75cm 0 0 0}, clip]{spec_phot_mass_comp.pdf}
    \caption{Our chosen photometric mass estimation for the total stellar mass of a galaxy in SDSS compared to {\tt Firefly} total galactic stellar mass measurements using spectroscopy. We compare total stellar masses of the galaxies, as opposed to bulge or black hole mass estimates, as this measurement had the largest set of spectroscopic estimates available. Our methodology tends to systematically overestimate the total stellar masses for lower-mass galaxies but underestimate the total stellar mass for higher-mass galaxies. This systematic over and under estimation is seen through the points deviations from a one-to-one line, as shown in the plot on the right. Still, as a whole, our methodology is not overly far off from the `true' masses as the histogram that compares the difference in spectroscopic and photometrically estimated total galactic stellar mass has a mean of $\sim-0.14$, as shown by the figure on the left. There is, however, a rather large scatter in the difference, which has a standard deviation of $\sim0.47$. Left: A histogram of the difference in log space of the spectroscopic and photometrically derived total galactic stellar masses. Right: A  plot in log space of the spectroscopic versus photometrically calculated total galactic stellar masses. The dashed line is the one-to-one line. The contours enclose $5\%$, $20\%$,  $40\%$, $60\%$, $80\%$, and $90\%$ of the data.}
    \label{fig:mass_comp}
\end{figure*}

Our process adopts a flat mass-to-luminosity relationship to estimate galaxy stellar mass instead of a more fine-tuned one. Mass-to-luminosity relationships in the infrared are sensitive to certain galaxy parameters such as star formation rates \citep{leroyetal2019} and colour \citep{Culver_2014}. However, a colour dependent mass-to-luminosity relationship requires more than one reliable magnitude, and only some WISE band passes have consistently reliable measurements for a large number of objects. For instance, $W4$ measurements are generally noisy and shallower \citep{WISE}. Thus, as this study is attempting to be as inclusive of sources as possible, we adopted a constant mass-to-luminosity ratio. This practice tends to overestimate stellar mass for low-mass galaxies and underestimate the stellar mass of higher-mass galaxies as seen in Figure~ \ref{fig:mass_comp}. 

To find the black hole mass hosted by each source, we used the M$_{\textrm{BH}}$-M$_{\textrm{bulge}}$ relation from \citet{mcconnell2013}, which is derived from a suite of  black hole - host relations. Although these relationships have traditionally been obtained from higher mass objects \citep{mcconnell2013}, it has been shown that objects on the low-mass end follow a similar relationship, albeit with large scatter \citep{schutte19, baldassare20}. This assumption further holds as LISA MBHB host galaxies should have enough time to return to expected scaling relations after merger, which takes on the order of $\sim1$~Gyr \citep{Medling_2015_return_to_relation_timescale}, before the MBHB successfully coalesces, which could take $>1$ Gyrs \citep{Holley_Bockelmann_2015_bhs_merge_timescale_long, Khan_2018_bhswillmerge}. 

To use the M$_{\textrm{BH}}$-M$_{\textrm{bulge}}$ relation, we determined the fraction of the stellar mass that is held in the bulge through the inclusion of a bulge fraction variable in the mass calculation. The bulge fraction can be determined well within a range that is dependent on the morphology of the galaxy or a set of other features \citep{peng_2010_gal_morpho_bd, Kimbrell_2023_bd_decon_in_dwarf}. As galaxy morphology is not reported by SDSS on large scales, we adopted bulge fractions that bracket likely values. We split the estimation process to adopt two assumptions, a bulge fraction of $0.01$, indicating essentially no bulge, and a bulge fraction of $1$, indicating the galaxy is all bulge as would be seen in an elliptical galaxy. 

The black hole mass is then calculated using the \citet{mcconnell2013} relationship, 

\begin{equation}
    \textrm{log}_{10}\left(\frac{M_{\textrm{BH}}}{M_{\odot}}\right)=(8.46\pm0.08)+(1.05\pm0.11)\times\textrm{log}_{10}\left(\frac{M_{\textrm{bulge}}}{10^{11}M_{\odot}}\right).
\end{equation}

We used a Monte Carlo technique to estimate the allowed range of possible black hole masses hosted by each system included in the sky localisation cap. We assumed Gaussian error distributions for $W1$, z, and the coefficients in the \citet{mcconnell2013} relation and sampled them $100$ times to obtain a mass distribution using the above procedure. Due to the splitting along bulge fraction, we tracked two separate distributions per source in this error estimation process. For our final range of possible MBHB total mass, we pull the lowest mass estimate ($16^{th}$ percentile) from the $0.01$ assumption and the highest mass estimate ($84^{th}$ percentile) from the $1$ assumption. As a final step in the error estimation, we added the intrinsic scatter of $0.39$~dex \citep[][]{mcconnell2013} by adjusting the final percentile masses up or down by $0.39$~dex. The addition of this scatter ensured that we derived the highest possible black hole mass and the lowest possible black hole mass providing as inclusive of a range as possible. Any objects with a mass range estimate that did not agree are excluded from the census when including this constraint. 

It should be noted that we adopted the fiducial MBHB total mass used in the analytical fits as the true total mass for the simulated system with no uncertainties. Although it is possible to solve for the LISA determined uncertainties on the MBHB system's intrinsic total mass, the uncertainty derived from the photometric mass estimation methods far outweigh these. In this work, we found differences between the $16^{th}$ percentile and $84^{th}$ percentile estimations on the photometric masses to be as large as $3$~dex. Comparatively, LISA derived fractional errors on the intrinsic total system mass are expected to be at sub-percent levels for this mass and redshift space \citep{colpi2024lisadefinitionstudyreport}. 
 
\subsection{Framework Example}
\label{subsec: framework example}

\begin{figure*}
    \centering
    \includegraphics[width=0.75\textwidth]{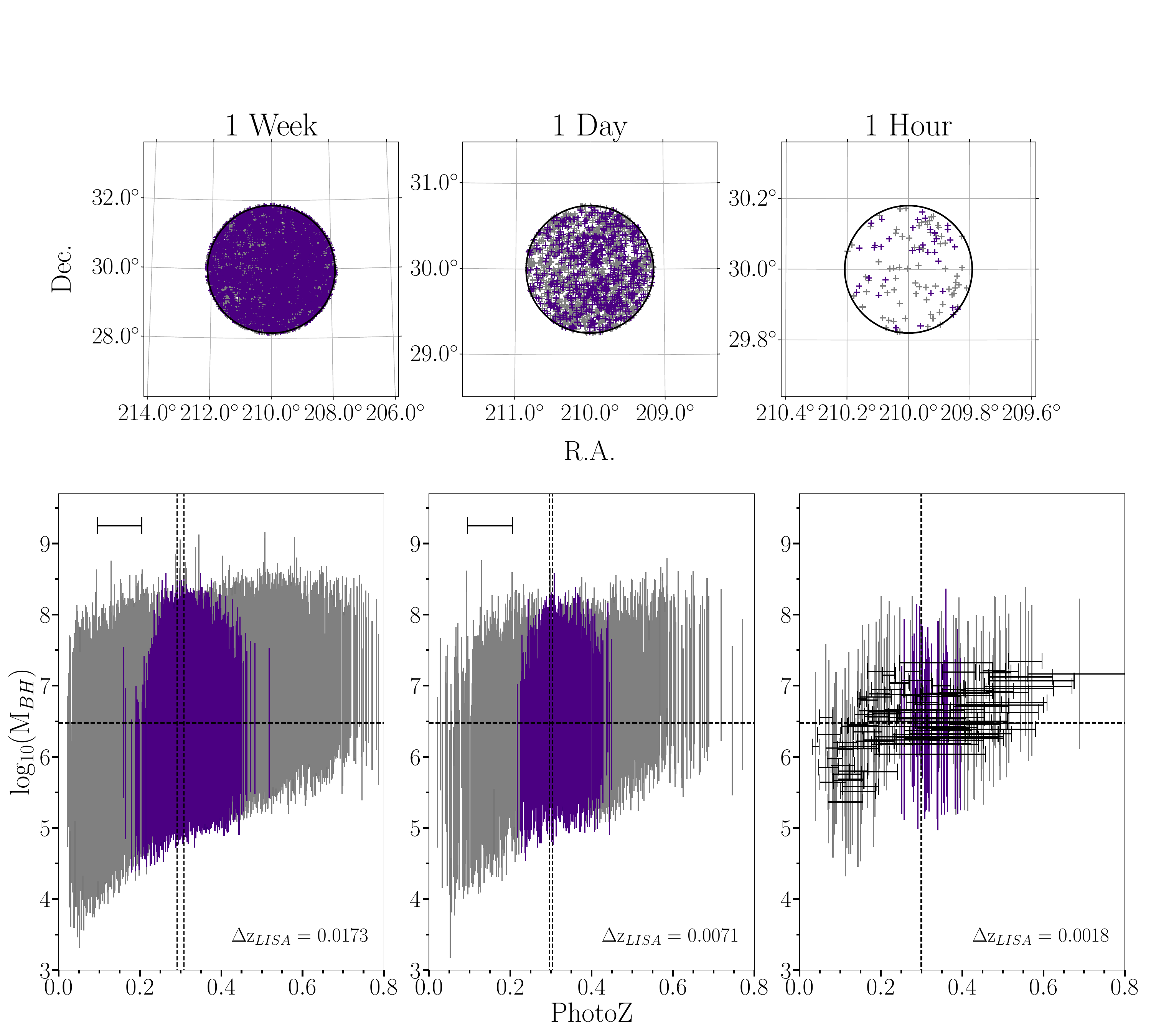}
    \caption{From left to right, framework progression from one week to one hour from coalescence when considering all median constraints for a $3\times10^{6}$~M$_{\odot}$ object at $z=0.3$. In all plots, objects plotted with dark purple are candidate host galaxies while objects in grey are galaxies that fail to meet our definition of a candidate host galaxy. Top row: Evolution of the sky localisation cap, represented by a thick black outline, for the simulated LISA system. The sky cap is plotted onto a zoomed-in Mollweide projection. Bottom row: Evolution of the census due to changing luminosity distance and mass constraints. The horizontal dashed line is the total mass used to simulate the LISA parameter estimations, in this case $3\times10^{6}$~M$_{\odot}$. The vertical dashed lines show the accepted range of redshift based on the LISA uncertainties. This range is stated explicitly in the lower right hand corner of each plot. The estimated range of the total mass of a black hole that would be hosted by each source included in the LISA sky localisation region is represented by a bar with vertical extent. The typical error in redshift space for the sources in each sky cap is represented by a horizontal error bar in the upper left corner of the plot for all but one hour from coalescence. In the bottom right plot, the error bars are plotted onto the points as there are few enough for readability. For all bottom plots, if a galaxy's estimated total black hole mass intersects with the horizontal mass line and has a redshift that agrees within error of the boxed region of luminosity distance uncertainties, the galaxy is plotted in purple and is included in our final census. If a galaxy does not meet these criteria it is plotted as grey and is not included in our final census count.}
    \label{fig:indepth constraint plots}
\end{figure*}

Figure~\ref{fig:indepth constraint plots} provides an in-depth look at how our framework combines different constraints into one census. Specifically, Figure~\ref{fig:indepth constraint plots} showcases the framework's step-by-step functions for a simulated binary of $3\times10^{6}$~M$_{\odot}$ total mass at $z=0.3$ for time steps of $1$ week, $1$ day, and $1$ hour from coalescence and how sources in SDSS move in or out of the census based on the constraints simulated for LISA. 

Beyond showcasing the modular nature of this framework by depicting various combinations of constraints at differing time steps Figure~\ref{fig:indepth constraint plots} also illustrates how we sample SDSS data for an individual MBHB in sky, redshift, and mass space. Although the curated dataset is well sampled on a survey-wide scale, this is less true when increasing constraints from LISA approach the limits of the SDSS population. This increasing sparseness is evident one hour from coalescence when the number of sources on the sky is noticeably less in redshift and mass space compared to the two prior time steps, pointing to the fact that SDSS is not a deep enough survey to fully characterize the desired mass and redshift space close to coalescence.

\section{A Census of Select Simulated LISA Systems}
\label{sec:Results}
We used our framework to conduct a census of candidate host galaxies galaxies for a variety of simulated MBHB systems with differing properties or sky locations. These iterations illustrate how changing the expected system properties impacts the number of galaxies we can expect to find when searching for an EM counterpart. 

Following \citet{mangiagli20}, we considered MBHBs of three total masses: heavy ($1\times10^{7}$ M$_{\odot}$), intermediate ($3\times10^{6}$ M$_{\odot}$) and light ($3\times10^{5}$ M$_{\odot}$). The intermediate system mass is notable as it sits in the most sensitive part of the LISA sensitivity curve. Additionally, for the intermediate system, we simulated two different redshift depths, $0.3$ and $0.5$. Lastly, we varied the central position of a simulated system of constant total mass, redshift, and time until coalescence between $100$ different randomized sky positions in the SDSS survey footprint. For the census, we considered each constraint (i.e. sky position, luminosity distance, mass) individually to understand its influence and implications. Then, we combined all three constraints for a final, best-case census. In most cases, we investigated the time steps of $1$ month, $1$ week, $1$ day, $1$ hour until coalescence, as well as the parameter estimates calculated after coalescence.

\subsection{Impacts of the Total Mass of the Binary }
\label{subsec: changing mass}

We began the census exercise by counting the candidate host galaxies in SDSS for the three separate mass regimes at a fixed redshift of $0.3$ considering various combinations of constraints at all confidence levels provided by the \citet{mangiagli20} fits. For the heavy mass regime, at times further from coalescence, the query we designed was too large to execute, so we only conducted a census starting at the $1$ day time step, or if only the median constraints were considered, starting at the $1$ week time step. 

\begin{figure*}
	\centering
	\includegraphics[width=\textwidth]{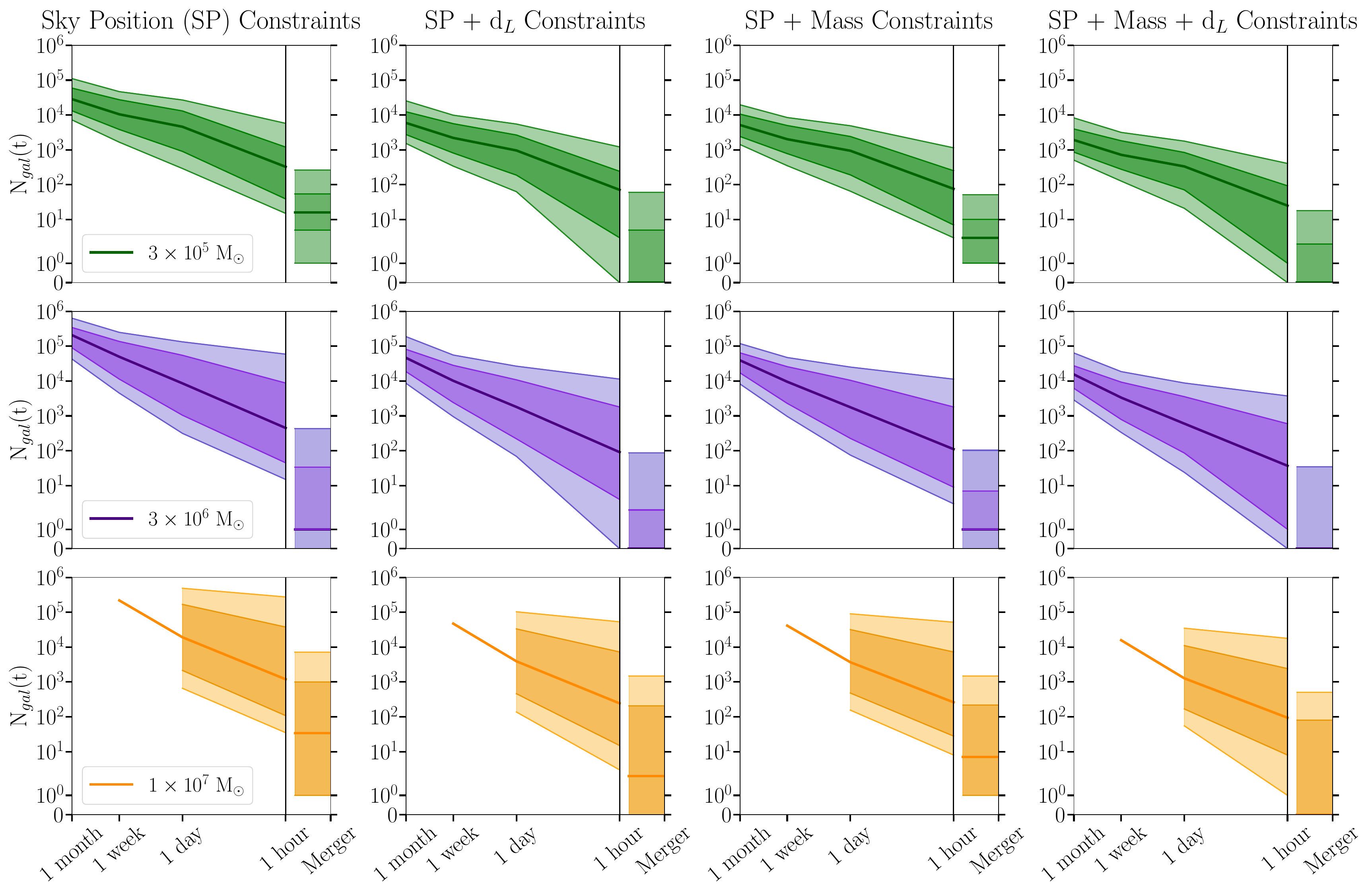}
    \caption{From top to bottom, constraints as applied to the SDSS survey for the light,  intermediate, and heavy masses at (RA, Dec)$=$($210$, $30$) and $z=0.3$ from one month from coalescence through one hour from coalescence as well as after coalescence. The solid line is the number of candidate host galaxies that agree with the median level of constraints, with the lighter shaded region being the number of candidate host galaxies that agree with the upper and lower $68^{th}$ percentile confidence level and the lightest shaded region being the number of candidate host galaxies that fall within the upper and lower $95^{th}$ percentile confidence level.}
    \label{fig:waterfall plots}
\end{figure*}

The results of the census for the three different total mass systems are shown in Figure~\ref{fig:waterfall plots}. Generally, we found that as mass increases, the number of candidate host galaxies, no matter the constraints used, also increased. One exception is that, beginning around $1$ day from coalescence, the light and intermediate systems census numbers begin to converge. In fact, at $1$ hour when considering all median constraints, there is only a difference of $12$ sources between these two different total mass systems. There is an apparent drop to zero sources for both light and intermediate sources when considering the lower $95^{th}$ percentile, or most stringent, constraints, which is seen again after coalescence for both systems even at the median constraint level. The seen drop to zero sources is not a physical phenomenon but is due to the completeness of SDSS. Briefly, it is not that the constraints have effectively culled all candidate host galaxies, but that SDSS has a limiting magnitude which effects the underlying statistics of our source population. Furthermore, while it appears that in Figure~\ref{fig:waterfall plots} both the mass and luminosity constraints affect the total source count similarly when paired with the sky localisation area, this is not entirely true. Due to the lack of robust photometric mass estimates, the luminosity distance constraint acts as a more informative constraint. We discuss the implications of these behaviours in Section~\ref{section: discussion}. 

\begin{figure}
    \centering
	\includegraphics[width=0.75\columnwidth]{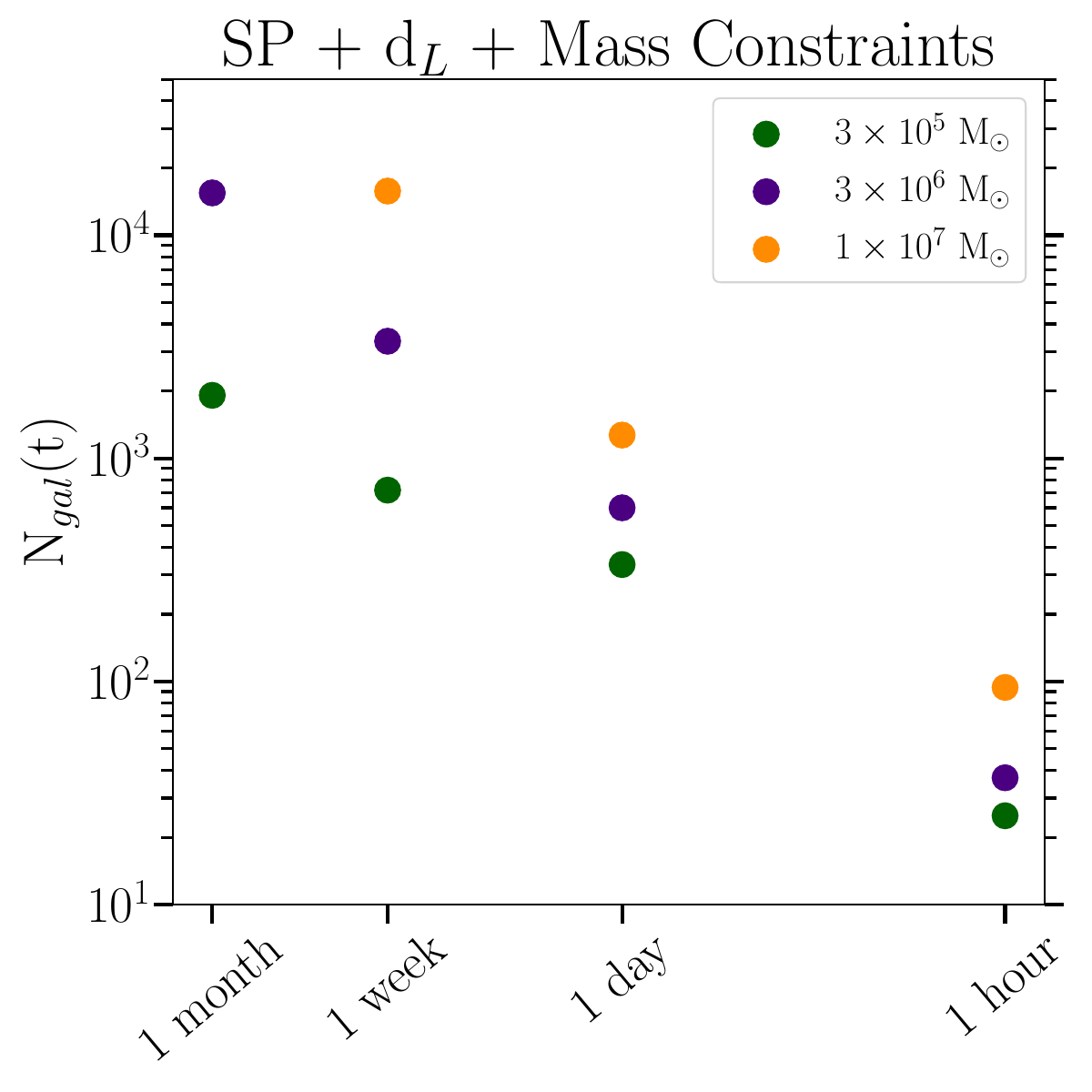}
    \caption{Generally, when considering median constraints, all three mass ranges have census numbers that evolve similarly in time and end with a range of $10$s$-100$s of objects one hour from coalescence. The points on this plot represent the median census numbers taken for light, medium, and heavy weight systems at a redshift of $0.3$ and an (RA, Dec)=($210$, $30$). For the heavy system, there is no data point at one month from coalescence as that time was unable to be measured for this weight of system.}
    \label{fig:med counts}
\end{figure}

 The intermediate mass regime has the most steady decline while both the light and the heavy mass regimes have some areas of differing evolution with time. These changes in evolution are most clearly depicted in Figure~\ref{fig:med counts}. However, these areas of differing evolution are in agreement with the depicted analytical fits to the parameter uncertainties of \citet{mangiagli20} (see their Figures $7-10$ for reference).

\subsection{Impacts of Binary Redshift}
\label{subsec:changing red}
We next performed a census focusing on how varying a fiducial system's redshift affects the final source count. 

\begin{figure*}
    \centering
    \includegraphics[width=\textwidth]{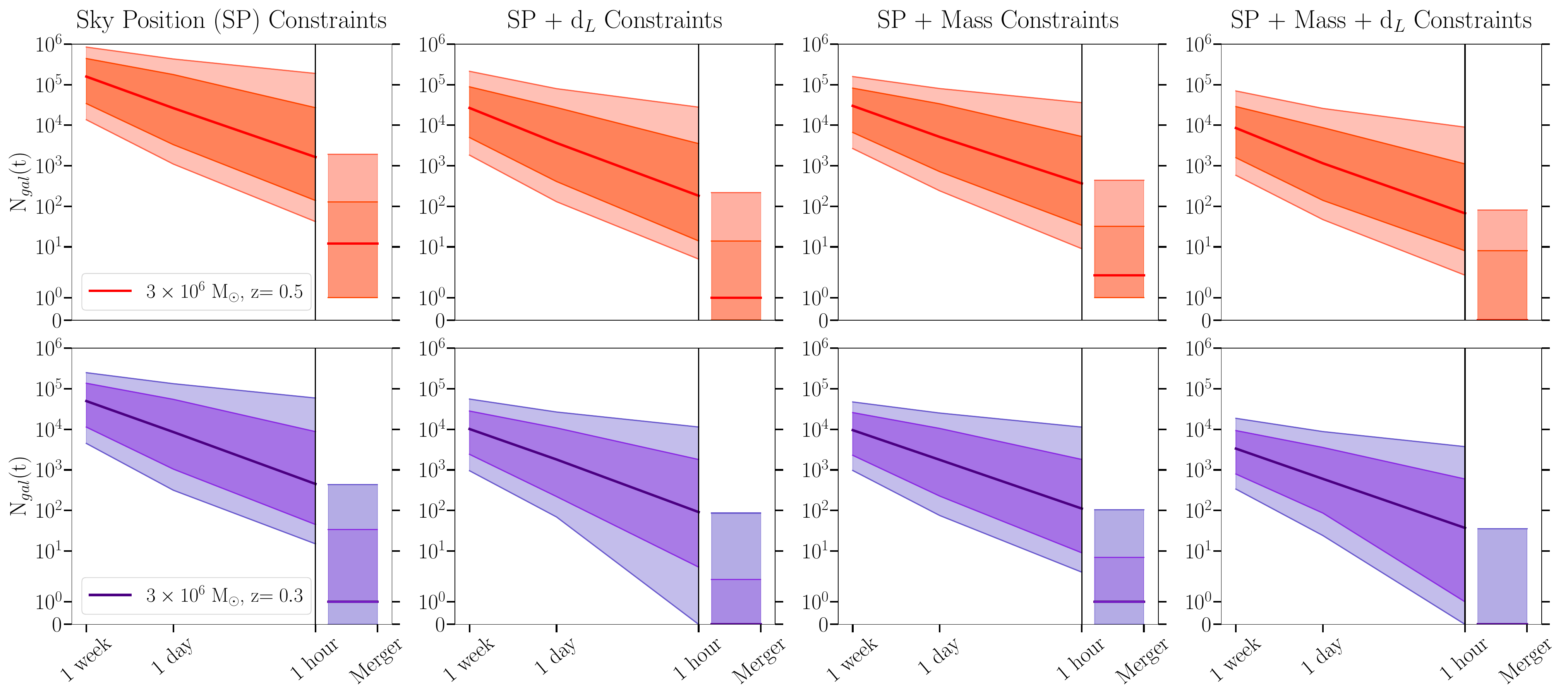}
    \caption{Similar to Figure~\ref{fig:waterfall plots} but instead displaying the effects of changing redshift. Mass is kept constant at $3\times10^{6}$M$_{\odot}$ while the redshift is varied between $0.3$ and $0.5$ for all constraint combinations. Although the median count begins slightly higher, all census counts tend to evolve slower with time and be less influenced by additional constraints at higher redshifts when compared to the census counts seen at lower redshifts.}
    \label{fig:waterfall plots change z}
\end{figure*}

To perform this test, we simulated a system with a total mass of $3\times10^{6}$~M$_{\odot}$ at two separate redshifts, $0.3$ and $0.5$. We then conducted a census of candidate host galaxies for all confidence levels from 1 week until coalescence. The results are presented in Figure~\ref{fig:waterfall plots change z}. We note the change in number of time steps investigated in this portion of our work due to query limitations. 

Overall, the higher redshift resulted in census numbers that are about a factor of $3$ higher than the lower redshift at the same time step. For instance, the number of candidate host galaxies in the median sky localisation constraint at $z=0.5$ one week from coalescence is $\sim150,000$ compared to $\sim50,000$ for $z=0.3$.  This difference in source counts is a result of the interplay of two different effects. Firstly, at higher redshifts the constraints on all uncertainties become less strict for the same total mass system \citep{mangiagli20}. The higher redshift constraints do, however, evolve similarly to the lower redshift constraints through time. The similarity of the constraint evolution in time leads to the overall similar evolution of the census curves in Figure~\ref{fig:waterfall plots change z}, though the higher redshift census does have a slightly less steep evolution. The similar evolution of census counts over time can be seen most clearly when comparing the final plots. Secondly, even for identical parameter estimates, having systems at a higher redshift inherently leads to probing bigger volumes. It should be noted that there is a higher incidence of galaxies in the curated dataset at a redshift of $0.3$ compared to $0.5$, meaning that more sources being in agreement at a higher redshift may be even more pronounced when a census is taken for a deeper survey.
 


\subsection{Impacts of the Sky Position of the Binary}

We investigated the effects of varying the central position of our simulated MBHB system throughout the central portion of the SDSS survey footprint to better characterize the distribution of candidate host galaxies throughout the survey. This exercise captures the impact of any anisotropies in the observed data distribution that exist naturally in large scale surveys. 

To probe how important these effects are for our census, we choose at random $100$ pointings within the survey footprint that are equally distributed in area on the sky. We then applied the sky localisation cap based on the simulated median uncertainties assuming a $3\times10^{6}$~M$_{\odot}$ MBHB at $z=0.3$ an hour until coalescence. One hour from coalescence was chosen as it prevents the double counting of sources by ensuring that the caps do not overlap at any point, while still fitting many different caps inside the portion of the SDSS footprint that has dense and relatively uniform observational coverage. 

\begin{figure}
	\includegraphics[width=\columnwidth]{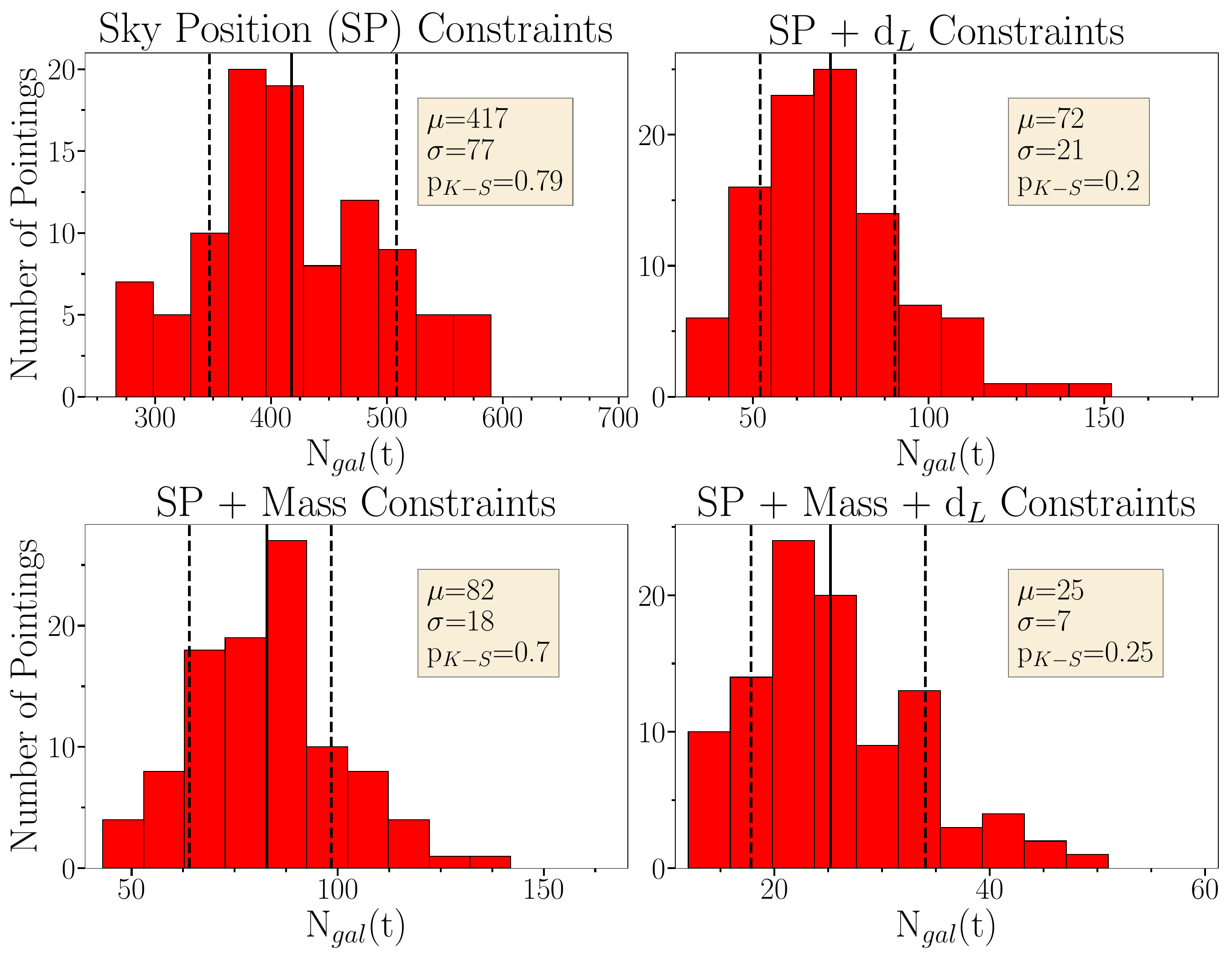}
    \caption{The census of objects based on 100 random RA and Dec positions in the SDSS footprint. All constraints were based on the median level fit and implemented the intermediate mass, at $z=0.3$, and $1$ hour from coalescence. The x-axis shows the number of candidate host galaxies and the y-axis depicts the number of pointings which contain that number of galaxies. The solid line denotes the median value of each distribution and the dashed lines denote the $16^{th}$ and $84^{th}$ percentile value of each distribution.}
    \label{fig:sp hists}
\end{figure}

The results of this exercise are illustrated in Figure~\ref{fig:sp hists}.  Half of the histograms in Figure~\ref{fig:sp hists} strongly follow an underlying normal distribution, namely those on the left hand side. The fact that these distributions are near normal is supported by a Kolmogorov-Smirnov (K-S) test. The K-S test quantifies the degree to which a particular distribution's cumulative density function (CDF) deviates from the CDF of a normal distribution. The more deviation between the CDF's the less likely it is that the distribution of interest is normal in nature. The amount of this deviation is quantified by a statistic called a p-value. The distributions on the left hand side of  Figure~\ref{fig:sp hists} have a p-value of $\geq0.5$, indicating that there is insufficient evidence to claim a non-normal distribution. However, when the luminosity distance constraint is included, the significance of normality for the distributions in Figure~\ref{fig:sp hists} decreases as evidenced by a drop in p-values from $\geq0.5$ to $\sim0.2$. The mean and standard deviation values for each distribution are listed in each respective plot. Based on these standard deviations, Figure~\ref{fig:sp hists} shows that varying the sky position can differ the final source count of our census by at most a few tens of sources, especially in the case where all constraints are considered.


\section{Discussion}
\label{section: discussion}

\subsection{Comparisons to Previous Work}
\label{comparisons}
In general, we found that at $1$ hour until coalescence, the final source count ranged from tens to thousands of objects depending on the simulated MBHB system. And after coalescence, for systems of all masses, when considering all median constraints, no candidate host galaxies stay in our census, seen in the furthest right hand column in Figure~\ref{fig:waterfall plots}.  As mentioned in \S\ref{sec: intro}, previous census work has been performed by a number of groups; unfortunately, due to updated LISA configurations, as well as a difference in techniques and investigated mass and redshift spaces, a direct comparison between our work and much of the previous work becomes difficult \citep{Holz_2005, Kocsis_2008}. 

However, \citet{lopsetal2023} not only also investigated simulated galaxies, but performed census work at a very similar mass and redshift space to what is investigated in this work. A more direct comparison is thus warranted. \citet{lopsetal2023} found that, for a $3\times10^{6}$M$_{\odot}$ system at $z=0.3$ within $10$ hours until coalescence, between $5$-$200$ sources were within the $90^{th}$ percentile of agreement with their constraints. In contrast, our work employed a hard constraint cut opposed to a probability cut and did not look explicitly at $10$ hours prior to coalescence. Nevertheless, our numbers fall within the same range for the same total binary mass and redshift. Thus, investigations of the type performed in this paper using EM observations are not necessarily at odds with complementary investigations using cosmological simulations and do not suffer major discrepancies. Detailed comparisons between these two approaches are likely to be a powerful approach to probing the mitigating limitations of each technique.

\subsection{Limitations}
\label{limitations}

This framework is intended to be a first step towards the coordination of coincident GW and EM signals with LISA. However, there are some limitations to our implemented methodology which fall into categories related to the simulated binary, the LISA parameter estimation techniques, and the treatment of the real EM sources.

\subsubsection{LISA Limitations}
Currently, our framework relies on analytical fits for the parameter estimates where the total binary mass is chosen as an input. Our framework, then, treats the total binary mass as a known parameter without uncertainty. This scenario fails to capture the realism of an on-the-fly detection where LISA will provide a range of total binary mass estimates and not a singular static value. As a result, this framework does not yet adopt a realistic mass treatment using LISA constraints for chirp mass and mass ratio as a means to estimate total binary mass or the translation between the binary system's rest- and detector-frame mass. 

Furthermore, our framework assumes that the LISA sky localisation region is circular. In reality, LISA's sky localisation area may not be continuous, having non-contiguous areas on the sky that show equal probability for being the true host's location \citep[see for example Figure 6 in][]{Katz_2020}. Including a more accurate localisation footprint is important for developing observational strategies to better characterize possible difficulties with natural fluctuations in the survey and other observational constraints such as slewing time (i.e. retargeting time). Additionally, the analytical fits we use overestimate the uncertainties occasionally, especially in the lower mass regime close to the time of coalescence for our redshift space (see Discussion in \citealt{mangiagli20}). This characteristic of the fits along with the unrealistic sky localisation footprint again leads to the more statistical nature of our analysis and leads us to de-emphasize the exact numbers found through our census work, instead focusing on orders of magnitude of source counts.

Finally, a more sophisticated approach to LISA parameter estimations would take into account additional errors on the luminosity distance due to weak lensing and peculiar velocity effects. Weak lensing occurs due to the inhomogeneous amount of matter GWs encounter between the source and the observer and leads to the distortion of the luminosity distance measurement through the deflection of GWs \citep{Holz_2005, Kocsis_2006, mangiagli2025}. Peculiar velocities describe the components of an object's velocity that are due to sources outside of the Hubble flow and impacts how accurately cosmological effects are taken into account when estimating luminosity distance \citep{Kocsis_2006, mangiagli2025}. Both weak lensing and peculiar velocity have a non-negligible effect on the uncertainty estimation for luminosity distance from GWs at low redshifts, although peculiar velocity errors are thought to be predominant over weak lensing errors \citep{Kocsis_2006, mangiagli2025}. As these sources of error are not taken into account in the analytical fits used in the framework, our luminosity distance uncertainties from GWs are likely underestimated.

\subsubsection{EM Limitations}
We made a variety of assumptions relating to the EM aspect of this work. Firstly, we assumed that each galaxy observed by SDSS is a non-active galaxy which may affect our used K-corrections, as they are reliant on the underlying spectrum of the source \citep{hogg2002k}. 

Secondly, we did not gather any morphology information from SDSS. The available morphology information, while inclusive of a relatively large sample set of galaxies across many previous works \citep{simardetal2011, mendeletal2014, bottrelletal2019}, still fails to capture the number of sources that are included in our curated dataset. These catalogues often fail to reach the redshift range that is investigated in our work due to the low photometric observation resolutions at these redshifts in SDSS \citep{simardetal2011}. This lack of morphology information makes the mass calculation very uncertain as morphology has a direct effect on the mass-to-light ratio through the bulge fraction. We attempted to compensate for this effect in our work by including a range of bulge fractions that would be possible for a dwarf galaxy, specifically a bulge fraction range of $0.01 - 1$. 

\begin{figure}
    \centering
    \includegraphics[width=0.85\linewidth]{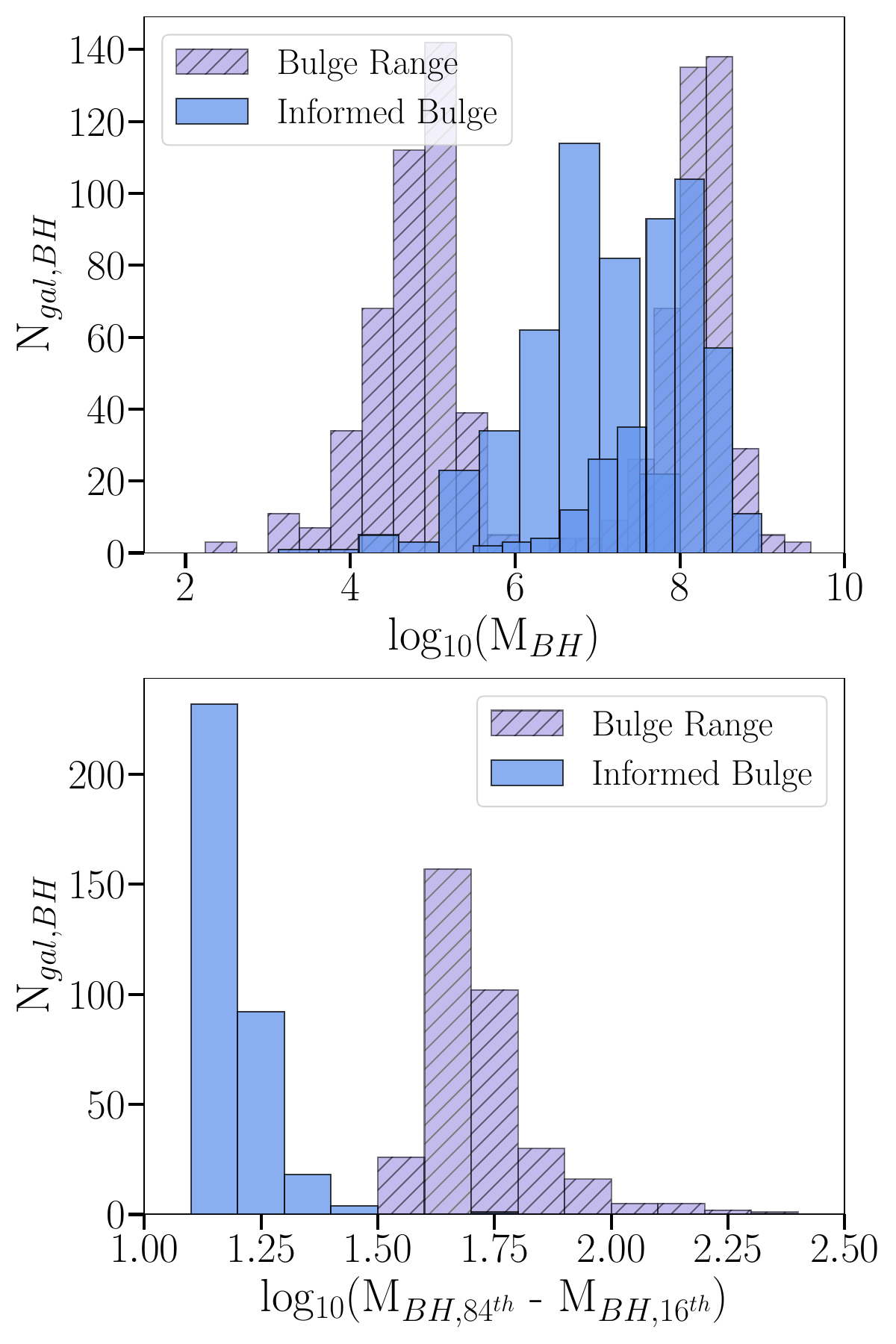}
    \caption{The differences in the range of estimated black hole mass using our methodology with an unconstrained range of bulge fractions compared to our methodology using estimated bulge fractions from \citet{simardetal2011} for a select set of galaxies. Black hole masses estimated with our original unconstrained bulge fractions tend to overestimate on the low mass end and have a total range of estimated black hole mass that is $\sim0.5$~dex more than what is found using an informed bulge fraction estimate. Top: Distributions of the $84^{th}$ and $16^{th}$ percentile estimates of black hole total mass found for each galaxy. The distributions found using the unconstrained bulge fraction range are denoted by a crosshatch styling. Bottom: Distributions of the difference between the $84^{th}$ and $16^{th}$ percentile estimates for total black hole masses that would be hosted by each galaxy. Using an informed bulge fraction range in our methodology creates a distribution with a mean of $\sim1.17$ and a standard deviation of $\sim0.06$. Using the unconstrained bulge fraction range in our methodology creates a distribution with a mean of $\sim1.69$ and a standard deviation of $\sim0.15$ and is again denoted by a cross hatch styling.}
    \label{fig:bulge comp}
\end{figure}

To probe how our lack of bulge fraction information for each candidate host affected our photometric mass estimates of the central black holes, we compared our original mass estimation methodology against a select sample of $421$ galaxies from \citet{simardetal2011} that have bulge fraction estimates and are in our curated dataset. For the first step in this comparison, we calculated the estimated black hole mass for these galaxies with known bulge fractions assuming the $0.01 - 1$, or unconstrained, bulge fraction range. We then performed the same calculation, but instead limited the bulge distribution to a normal distribution with mean and standard deviation based on the bulge fraction value and error reported by \citet{simardetal2011}. The results of this comparison are shown in Figure~\ref{fig:bulge comp}.

Generally, having bulge fraction information results in an estimated black hole mass range that is $\sim0.5$~dex more constrained and with $\sim0.1$~dex less variance in the distribution when considering the $68^{th}$ percentile than what is found without bulge fraction information for this set of galaxies. Additionally, we see that the black hole masses found using the unconstrained bulge fractions are overly inclusive on the low-mass end compared to what is found using the informed bulge fractions. The low-mass range of black hole masses is what LISA will be most sensitive to, so the inclusion of bulge information will most likely make for a much more informative mass constraint in the future.

It should be noted that, due to the redshift range of the catalogue used in this exercise, this set of sample galaxies are at a much lower redshift than what will most likely be detected by LISA. The mean redshift of the galaxies used in this comparison is $z\simeq0.1$ and only $\sim6\%$ have a $z~\ge0.2$. Because these sample galaxies are at a lower redshift, we would expect these sample galaxies to be biased towards a higher black hole mass compared to the candidate host galaxies at higher redshifts more aligned with LISA's capabilities based on the mass function of galaxies. However, since we also expect these candidate host galaxies to be generally less luminous and bluer, they should be fainter in the infrared, leading to fewer detections at high redshifts in surveys like WISE. The interplay of the observational limitations that arise with changing redshift and how they will affect detections of candidate host galaxies is difficult to characterize with observational data alone and points to the importance of complementary simulation work to help characterize the limitations arising from an unconstrained bulge fraction.

Lastly, we chose not to include a fine tuned mass-to-light ratio as the $(W1-W4)$ colour is unreliable due to generally noisy $W4$ data \citep{WISE}. The last two choices lead to an inclusive but not conservative census, and in fact our calculated photometric mass for the binaries hosted by each galaxy span $\sim2.5$ orders of magnitude on average.

We compensated for each of the above limitations in a manner we felt conservative. For example, we focused on a mass and luminosity distance regime that is well constrained by the \citet{mangiagli20} analytical fits. We included a representative sky localisation area that is the same final total area on the sky although not necessarily the most accurate shape. And we developed an overly inclusive mass calculation methodology.

\subsection{Implications for Planning LISA Follow-up}
\label{implications}
The product of our work highlights important obstacles and nuances that will be relevant for developing LISA follow-up methodologies. We now discuss the implications of how changing mass, redshift, and sky position of the simulated system affect the census, and discuss in more depth some of the peculiarities of our census.


\begin{figure*}
    \centering
    \includegraphics[width=0.75\textwidth]{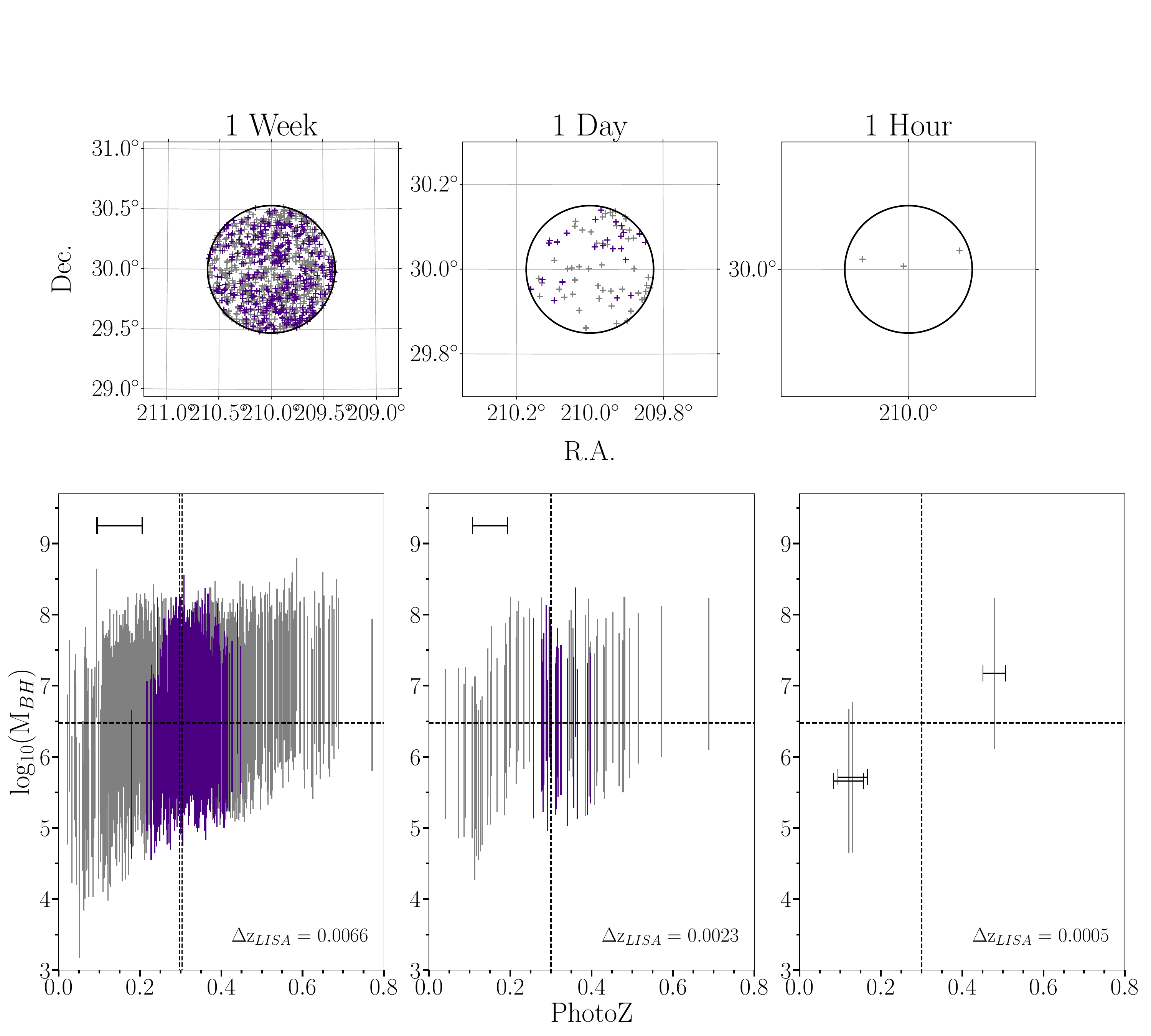}
    \caption{A more in-depth look as to the possible causes behind the drop out of objects for the lower $95^{th}$ percentile uncertainties in our census. Overall, it seems that SDSS is too sparsely sampled at this intersection of sky position, redshift, and mass space along with the fact that LISA parameters are most constrained at this level which causes the apparent absence of candidate host galaxies in our census. This plot follows Figure~\ref{fig:indepth constraint plots} exactly but for the lower $95^{th}$ percentile parameter estimations for the same simulated system.}
    \label{fig:indepth-constraint-plots-95}
\end{figure*}

The census and its evolution with time tell us about the factors that determine the candidate host galaxies. The evolution of all mass regimes follows a similar pattern of decreasing with time, in fact ending with very similar final source counts.  At times far from coalescence, this behaviour is mostly influenced by the evolving signal-to-noise curve of LISA. In other words, the beginning of the census is depicting the known behaviour that the longer a source is detected in the LISA band the tighter constrained the parameter estimations become on average (although this further constraining with time of the parameter estimates is not necessarily true when considering the confidence intervals, which may widen closer to coalescence, see Figure 7 of \citealt{mangiagli20}). It should be noted that all of our simulated systems are above the assumed LISA detection limit of signal-to-noise $\geq8$ at all confidence intervals and investigated times from coalescence \citep{mangiagli20}. However, beginning as early as a day from coalescence for some confidence intervals, the census counts are mainly influenced by the underlying dataset as there are so few possible sources (see Figure~\ref{fig:indepth constraint plots} and Figure~\ref{fig:indepth-constraint-plots-95}). For instance, although the mass function of black holes is weighted towards the low-mass end, larger black holes are generally hosted in larger, and thus brighter, galaxies. These larger galaxies, which would not be LISA host galaxies, are preferentially observed in a shallower survey such as SDSS. As our mass estimation is currently poorly constrained, these larger galaxies may inflate our census numbers until they are ruled out right before coalescence. 

Survey completeness has important implications for the census in general. One clear example is the apparent drop to zero sources one hour from coalescence for the intermediate mass system as seen in Figure~\ref{fig:waterfall plots}. This behaviour highlights the complex completeness concerns that arise when combining real source population statistics, LISA detection parameters, and EM observational constraints. The distribution of SDSS and WISE observations in our curated dataset that agrees with the sky localisation, luminosity distance, and mass constraints is sparse. This sparsity is best showcased by Figure~\ref{fig:indepth-constraint-plots-95} which depicts how the time-evolving LISA parameter estimates at the $95^{th}$ lower confidence interval for an intermediate mass system at $z=0.3$ interacts with the underlying statistics of the curated dataset. A result of this sparseness is that, as the LISA constraints tighten, there are not many sources with SDSS and WISE observations that agree with the LISA constraints. In other words, this drop to a very small number of sources is most likely due to a completeness issue. For instance, when compared to the COMBO-17 survey, SDSS may have a completeness as poor as $50\%$ at magnitude m$_{r, AB}\sim22$ for photometric galaxies\footnote{\url{https://classic.sdss.org/dr7/products/general/completeness.php}}.

\begin{figure}
    \centering
    \includegraphics[width=\linewidth]{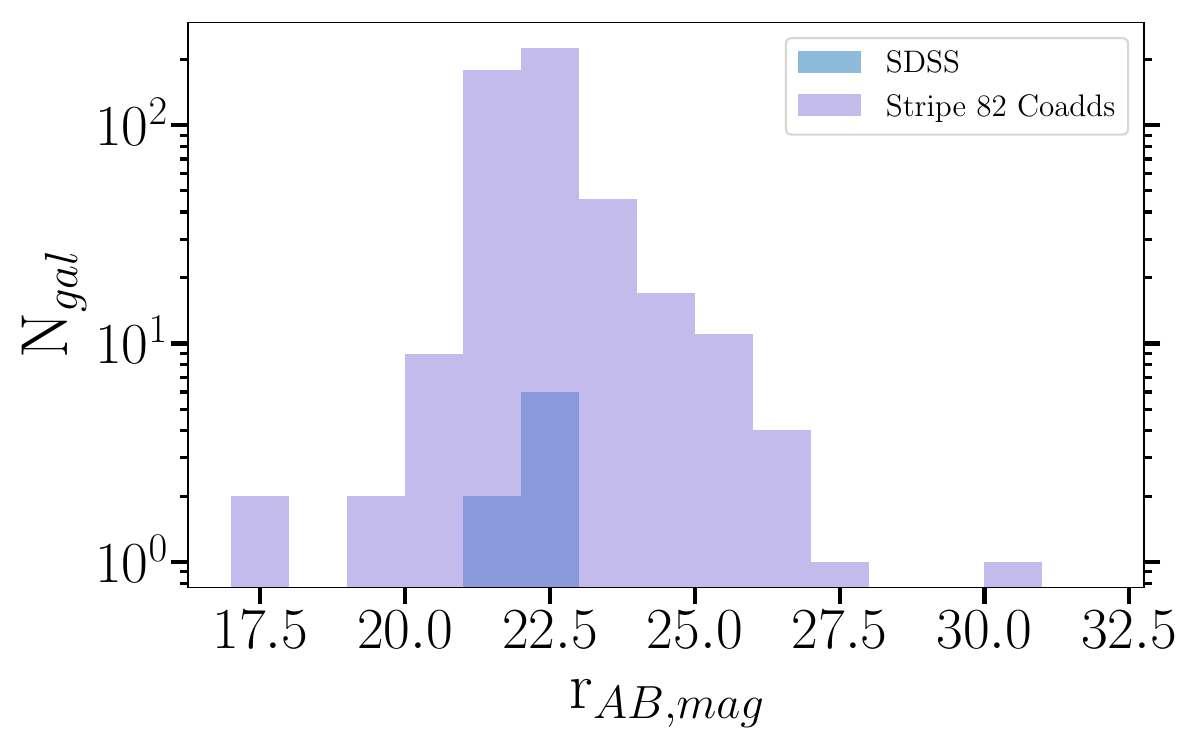}
    \caption{Comparison of census numbers when applied to the general SDSS footprint and the {\tt Stripe82} region of SDSS. In general, we see that the added depth of {\tt Stipe82} increase our candidate host galaxy counts by one to two orders of magnitude when only considering sky localisation regions from LISA. The increase in source counts points to the importance of survey depth limitations as we move forward with this work. Shown are the non-normal distributions of candidate host galaxies' $r$ magnitudes from SDSS (blue) and the {\tt Stripe82} coadds (purple). These distributions are based on median census numbers found using only the sky localisation constraint assuming an intermediate mass system at $z=0.3$ and coalescence. The y-axis is displayed using a log-scale.}
    \label{fig:stirpe82}
\end{figure}

To quantify the impact of completeness on the census, we applied our framework using only the median sky localisation constraints for a $3\times10^{6}$~M$_{\odot}$ system at $z=0.3$ at coalescence to the {\tt Stripe 82} region of SDSS. {\tt Stripe 82} spans $-50^{\circ}<$~RA~$<60^{\circ}$ and $-1.26^{\circ}<$~Dec~$<1.26^{\circ}$ and has been imaged at more epochs than the majority of the SDSS footprint, creating co-added photometric observations that are approximately two magnitudes deeper \citep{Annisetal14_stripe82, Jiang_2014_s82}. By applying our framework to this region, we illustrate how many faint objects are not observed in the main SDSS survey and are missed by our census. In {\tt Stripe82}, our census numbers increase by one to two orders of magnitude, seen in the non-normalized distributions in the left hand plot of Figure~\ref{fig:stirpe82}. Not only do we see a marked increase in the number of candidate host galaxies, but the census is generally less sparse and better characterizes the distribution of magnitudes. However, even these deeper co-added magnitudes from {\tt Stripe82}, fail to capture many candidate host galaxies. Simulations estimate that for masses between $10^{5}-10^{6}$~M$_{\odot}$ at $z=0.3$ less than $60\%$ of LISA MBHB host galaxies are captured by the co-added SDSS magnitude depth \citep{izetal2023}. Moving forward, we expect deeper surveys like LSST to help mitigate this completeness concern, but as LISA will be able to detect MBHB throughout the Universe, we will always be pushing against completeness limits. It is thus necessary to build tools that incorporate all aspects of what determines completeness including but not limited to observing conditions, cadence, and location on the sky. Fully characterizing this complex interplay of constraints and survey completeness is best done in conjunction with simulations, as simulations have a more complete knowledge of the parameter space associated with their data. 

The difficulty of measuring each source's bulge fraction introduces another completeness concern. Bulge fractions are indicative of the total brightness of the extended source. Given a galaxy with a small bulge fraction but a more massive disc leads to the galaxy having a comparatively brighter magnitude than if it had a larger bulge fraction and smaller disc. These changes in magnitude driven by bulge fractions are particularly notable for our work, which is influenced by survey limitations such as imaging depth. For instance, a MBHB system of mass $3\times10^{6}$~M$_{\odot}$ at a redshift of $z~=0.3$ with a bulge fraction of $0.01$ would be hosted in a galaxy that has a $W1$ magnitude of $\sim13$ while the same system with a bulge fraction of $1$ would have a host galaxy with $W1$ magnitude of $\sim18$. The former host galaxy would be detected in WISE (which has a limiting magnitude in $W1$ of $\sim16$ \citealt{WISE}), while the latter galaxy would not be detected and would be missing from our census.  

LISA will be able to detect systems at very high redshifts. However, the host galaxies of the MBHB systems that LISA will be most sensitive to, though, tend to be small and faint, abutting the flux limitations of many large-area sky surveys. To probe how these flux limitations affect our census numbers and whether it is feasible to expect to use the full breadth of LISA's capabilities, we applied our census framework to a higher redshift of $0.5$. As we move to higher redshifts, we found that there are consistently more candidate host galaxies in SDSS as compared to the lower redshift investigation of the same mass and time from coalescence. This difference in census counts as we probe higher redshifts is largely driven by the fact that LISA parameter estimates are less constrained at these high redshifts as is seen in the Figures in \citet{mangiagli20}. In our curated data set, objects are twice as likely to be found at a redshift of $0.3$ compared to $0.5$ (see Figure~\ref{fig:datasets_g}) implying that objects are preferentially found at $z=0.3$ in dataset. However, as the census numbers at $0.5$ are an order of magnitude more than those at $0.3$, this bias in our curated dataset seems to not be a contributing factor towards the differences in census numbers. To better characterize LISA's observational follow-up capabilities at higher redshifts, a combination of deeper surveys and comparisons to simulations should be pursued, as well as improving the LISA parameter estimate techniques.

The sampling of the survey footprint has implications for the observational load that we may expect when performing real-time coordinated LISA follow-up. Generally, we found a factor of $\sim2$ difference between source counts across pointings when considering all constraints, as shown in the final plot of Figure~\ref{fig:sp hists}. This general range in source counts provides a first-order constraint on the number of sources that need to be observed, and thus the observation time it would take, to successfully observe a LISA EM counterpart in a survey like SDSS. Furthermore, all of the histograms derived from the randomized pointings generally follow a normal distribution. Source counts consistently agreeing with a normal distribution across pointings implies that local overdensities or biases that occur naturally in surveys are not expected to cause a single outlier pointing with an unexpectedly large number of sources. Thus, for future LISA follow-up strategy development, applying LISA parameter estimations to surveys in a general manner should provide a strong foundation to characterize the expected observational load for that survey.

\section{Summary}
\label{section: summary}

In this paper, we developed a framework that is the first step towards coordinating successful EM follow-up for future LISA detections using ground-based observatories. This framework uses analytical fits to LISA parameter estimation curves provided by \citet{mangiagli20}, specifically those for sky localisation and luminosity distance, as well as a photometric mass cut to perform censuses of candidate host galaxies by comparing these fits to SDSS and WISE photometric data. We carried out this analysis for multiple time steps from coalescence, as well as after coalescence, varying total MBHB system masses, and differing redshift depths. We simulated light ($3\times10^{5}$~M$_{\odot}$), intermediate ($3\times10^{6}$~M$_{\odot}$), and heavy ($1\times10^{7}$~M$_{\odot}$) mass systems at a redshift of $0.3$ one month through one hour from coalescence, as well as after coalescence. Additionally, to further probe how inherent flux limitations of surveys affect our census work, we investigated the census of an intermediate mass system at a higher redshift of $0.5$ one week through one hour until coalescence.  Finally, we applied the framework to $100$ random central positions through out the densest region of the SDSS footprint to characterize how natural variances in the distribution of sources in SDSS affect the results of the census. 

We found that in our curated dataset of SDSS and WISE EM data, there are between tens and hundreds of candidate host galaxies one hour from coalescence which agrees with previous work using simulations. The final census count is most influenced by the sky localisation and the luminosity distance uncertainties, while the mass constraint has the smallest effect. The mass constraint is comparatively less influential due to the large ranges of acceptable masses that are found using our photometric black hole mass estimation methodology. In order to create a more effective mass constraint, it is important to develop a mass estimation method that is more sensitive to information such as morphology or colour. Final census counts also closely followed the LISA signal-to-noise curve evolution that develops as a source is detected for longer periods of time. For instance, higher mass systems are detected for the least amount of time prior to coalescence and thus have a lower signal-to-noise ratio in the LISA band compared to the light and intermediate mass systems, leading to larger uncertainties and thus more galaxies being included in the final census count.

Importantly, our work highlighted various nuances that should be considered during future follow-up strategy development. Specifically: 

\begin{itemize}
    \item Probing to a higher redshift of $0.5$ for the intermediate mass system, which would still be considered a `local' system compared to the redshifts LISA will access, increased the number of host candidates by about a factor of three. This increase in census count is caused by less constrained LISA parameters, as well as the fact that probing a higher redshift inherently leads to probing a larger volume of space. Furthermore, this higher census count may be more pronounced in a deeper survey as this redshift is less thoroughly sampled comparatively to $z= 0.3$ in the curated dataset. This difference in source counts depending on redshift highlights the fact that follow-up of LISA MBHB detections is likely to be done at lower redshifts due to EM limitations.
    \item Through randomly sampling the SDSS footprint, we found that changes in final census counts generally follow a near normal distribution. A majority of pointings for a $3\times10^{6}$~M$_{\odot}$ at $z=0.3$ have about a factor $\sim2$ difference in source counts when considering all parameter estimates one hour from coalescence. These numbers provide an initial estimate for the observational load one may expect when performing future follow-up searches.
    \item The photometric estimation of the total binary mass that would be hosted in each galaxy is currently not a robust constraint, having a smaller effect on final census numbers compared to sky localisation and luminosity distance. Improvements in photometric mass estimation with EM methods are expected to make the total binary mass constraint a more robust cut. 
\end{itemize} 

Collaborators are actively investigating the above nuances from a simulation based perspective. Their complementary results will provide further insight into the issues addressed in this paper. Their work will be presented in an upcoming companion paper.

Overall, with our framework, we have developed the building blocks for future LISA-EM multi-messenger detection methodology. We have probed census numbers considering simulated MBHB systems of varying masses, redshifts, and times from coalescence for multiple combinations of LISA parameter estimates. Our research has provided a rough estimate of the future observational load for follow-up strategies as well as begun to characterize the extent to which we will be able to fully utilize LISA's extensive redshift detection depth. Furthermore, our work has brought to light a variety of limitations that affect our census and pointed to ways that these may be alleviated in future work. Most notably, our work has highlighted a complex completeness concern. Characterizing completeness has been shown to be a uniquely difficult problem when combining LISA parameter estimations, EM observational constraints, and underlying population statistics of real sources. By comparing this work to simulations that more directly follow the methodology executed in this framework and applying this framework to deeper surveys, future work will more fully characterize how completeness impacts our census numbers.

\section*{Acknowledgments}

C.L.D., J.C.R., T.B., K.F., M.E., and K.S. acknowledge support by the National Aeronautics and Space Administration (NASA) under award No. $80$NSSC$22$K$0748$. C.L.D. further acknowledges support from the NASA FINESST grant no. 80NSSC24K1478 and the EMIT National Science Foundation NRT-2125764 grant. The work of A.S. was supported by the National Science Foundation MPS-Ascend Postdoctoral Research Fellowship under grant No. 2213288. All authors would like to thank the anonymous referee for their insightful comments regarding this manuscript.


\section*{Data Availability}

The SDSS DR16 data that are used in this research are available through the SDSS online SQL Query client. The Query client can be reached at \url{https://skyserver.sdss.org/dr16/en/tools/search/sql.aspx}. The FireFly value added catalogue can be downloaded at \url{https://www.sdss.org/dr18/data_access/value-added-catalogs/?vac_id=47}. Access to the Python code used to form and run the framework referenced throughout this paper is available upon request to the author at \href{mailto:carolyn.drake@vanderbilt.edu}{carolyn.drake@vanderbilt.edu}. 
 



\bibliographystyle{mnras}
\bibliography{bibliography} 

\begin{thebibliography}{}
\makeatletter
\relax
\def\mn@urlcharsother{\let\do\@makeother \do\$\do\&\do\#\do\^\do\_\do\%\do\~}
\def\mn@doi{\begingroup\mn@urlcharsother \@ifnextchar [ {\mn@doi@}
  {\mn@doi@[]}}
\def\mn@doi@[#1]#2{\def\@tempa{#1}\ifx\@tempa\@empty \href
  {http://dx.doi.org/#2} {doi:#2}\else \href {http://dx.doi.org/#2} {#1}\fi
  \endgroup}
\def\mn@eprint#1#2{\mn@eprint@#1:#2::\@nil}
\def\mn@eprint@arXiv#1{\href {http://arxiv.org/abs/#1} {{\tt arXiv:#1}}}
\def\mn@eprint@dblp#1{\href {http://dblp.uni-trier.de/rec/bibtex/#1.xml}
  {dblp:#1}}
\def\mn@eprint@#1:#2:#3:#4\@nil{\def\@tempa {#1}\def\@tempb {#2}\def\@tempc
  {#3}\ifx \@tempc \@empty \let \@tempc \@tempb \let \@tempb \@tempa \fi \ifx
  \@tempb \@empty \def\@tempb {arXiv}\fi \@ifundefined
  {mn@eprint@\@tempb}{\@tempb:\@tempc}{\expandafter \expandafter \csname
  mn@eprint@\@tempb\endcsname \expandafter{\@tempc}}}

\bibitem[\protect\citeauthoryear{Abbott et~al.,}{Abbott et~al.}{2016}]{LIGO}
Abbott B.,  et~al., 2016, \mn@doi [Physical Review Letters]
  {10.1103/physrevlett.116.221101}, 116

\bibitem[\protect\citeauthoryear{Abbott et~al.,}{Abbott et~al.}{2017a}]{BNS}
Abbott B.~P.,  et~al., 2017a, \mn@doi [Phys. Rev. Lett.]
  {10.1103/PhysRevLett.119.161101}, 119, 161101

\bibitem[\protect\citeauthoryear{Abbott et~al.,}{Abbott
  et~al.}{2017b}]{Abbott_2017}
Abbott B.~P.,  et~al., 2017b, \mn@doi [The Astrophysical Journal Letters]
  {10.3847/2041-8213/aa920c}, 848, L13

\bibitem[\protect\citeauthoryear{Aghanim et~al.,}{Aghanim
  et~al.}{2020}]{Planck18}
Aghanim N.,  et~al., 2020, \mn@doi [Astronomy &amp; Astrophysics]
  {10.1051/0004-6361/201833910}, 641, A6

\bibitem[\protect\citeauthoryear{Ahumada et~al.,}{Ahumada
  et~al.}{2020}]{SDSS_dr16}
Ahumada R.,  et~al., 2020, \mn@doi [The Astrophysical Journal Supplement
  Series] {10.3847/1538-4365/ab929e}, 249, 3

\bibitem[\protect\citeauthoryear{Amaro-Seoane et~al.,}{Amaro-Seoane
  et~al.}{2017}]{amaroseoane2017laser}
Amaro-Seoane P.,  et~al., 2017, Laser Interferometer Space Antenna (\mn@eprint
  {arXiv} {1702.00786})

\bibitem[\protect\citeauthoryear{Amaro-Seoane et~al.,}{Amaro-Seoane
  et~al.}{2023}]{Amaro_Seoane_2023}
Amaro-Seoane P.,  et~al., 2023, \mn@doi [Living Reviews in Relativity]
  {10.1007/s41114-022-00041-y}, 26

\bibitem[\protect\citeauthoryear{{Annis} et~al.,}{{Annis}
  et~al.}{2014}]{Annisetal14_stripe82}
{Annis} J.,  et~al., 2014, \mn@doi [\apj] {10.1088/0004-637X/794/2/120}, \href
  {https://ui.adsabs.harvard.edu/abs/2014ApJ...794..120A} {794, 120}

\bibitem[\protect\citeauthoryear{Arun, Mishra, Broeck, Iyer, Sathyaprakash  \&
  Sinha}{Arun et~al.}{2009}]{Arun_2009}
Arun K.~G.,  Mishra C.~K.,  Broeck C. V.~D.,  Iyer B.~R.,  Sathyaprakash B.~S.,
    Sinha S.,  2009, \mn@doi [Classical and Quantum Gravity]
  {10.1088/0264-9381/26/9/094021}, 26, 094021

\bibitem[\protect\citeauthoryear{Baldassare, Reines, Gallo  \&
  Greene}{Baldassare et~al.}{2017}]{Baldassare_2017_agnindwarfs}
Baldassare V.~F.,  Reines A.~E.,  Gallo E.,   Greene J.~E.,  2017, \mn@doi [The
  Astrophysical Journal] {10.3847/1538-4357/836/1/20}, 836, 20

\bibitem[\protect\citeauthoryear{{Baldassare}, {Dickey}, {Geha}  \&
  {Reines}}{{Baldassare} et~al.}{2020}]{baldassare20}
{Baldassare} V.~F.,  {Dickey} C.,  {Geha} M.,   {Reines} A.~E.,  2020, \mn@doi
  [\apjl] {10.3847/2041-8213/aba0c1}, \href
  {https://ui.adsabs.harvard.edu/abs/2020ApJ...898L...3B} {898, L3}

\bibitem[\protect\citeauthoryear{Beck, Dobos, Budavári, Szalay  \&
  Csabai}{Beck et~al.}{2016}]{becketal2016}
Beck R.,  Dobos L.,  Budavári T.,  Szalay A.~S.,   Csabai I.,  2016, \mn@doi
  [Monthly Notices of the Royal Astronomical Society] {10.1093/mnras/stw1009},
  460, 1371

\bibitem[\protect\citeauthoryear{{Begelman}, {Blandford}  \& {Rees}}{{Begelman}
  et~al.}{1980}]{begelmanetal_1980}
{Begelman} M.~C.,  {Blandford} R.~D.,   {Rees} M.~J.,  1980, \mn@doi [\nat]
  {10.1038/287307a0}, \href
  {https://ui.adsabs.harvard.edu/abs/1980Natur.287..307B} {287, 307}

\bibitem[\protect\citeauthoryear{{Berti}, {Buonanno}  \& {Will}}{{Berti}
  et~al.}{2005}]{bertietal2005}
{Berti} E.,  {Buonanno} A.,   {Will} C.~M.,  2005, \mn@doi [\prd]
  {10.1103/PhysRevD.71.084025}, \href
  {https://ui.adsabs.harvard.edu/abs/2005PhRvD..71h4025B} {71, 084025}

\bibitem[\protect\citeauthoryear{{Blanton}, {Kazin}, {Muna}, {Weaver}  \&
  {Price-Whelan}}{{Blanton} et~al.}{2011}]{blanton_2011_nsatlas}
{Blanton} M.~R.,  {Kazin} E.,  {Muna} D.,  {Weaver} B.~A.,   {Price-Whelan} A.,
   2011, \mn@doi [\aj] {10.1088/0004-6256/142/1/31}, \href
  {https://ui.adsabs.harvard.edu/abs/2011AJ....142...31B} {142, 31}

\bibitem[\protect\citeauthoryear{{Bode}, {Haas}, {Bogdanovi{\'c}}, {Laguna}  \&
  {Shoemaker}}{{Bode} et~al.}{2010}]{bodeetal2010}
{Bode} T.,  {Haas} R.,  {Bogdanovi{\'c}} T.,  {Laguna} P.,   {Shoemaker} D.,
  2010, \mn@doi [\apj] {10.1088/0004-637X/715/2/1117}, \href
  {https://ui.adsabs.harvard.edu/abs/2010ApJ...715.1117B} {715, 1117}

\bibitem[\protect\citeauthoryear{Bogdanović, Bode, Haas, Laguna  \&
  Shoemaker}{Bogdanović et~al.}{2011}]{Bogdanovic_2011}
Bogdanović T.,  Bode T.,  Haas R.,  Laguna P.,   Shoemaker D.,  2011, \mn@doi
  [Classical and Quantum Gravity] {10.1088/0264-9381/28/9/094020}, 28, 094020

\bibitem[\protect\citeauthoryear{{Bonetti}, {Sesana}, {Haardt}, {Barausse}  \&
  {Colpi}}{{Bonetti} et~al.}{2019}]{bonettietal_2019}
{Bonetti} M.,  {Sesana} A.,  {Haardt} F.,  {Barausse} E.,   {Colpi} M.,  2019,
  \mn@doi [\mnras] {10.1093/mnras/stz903}, \href
  {https://ui.adsabs.harvard.edu/abs/2019MNRAS.486.4044B} {486, 4044}

\bibitem[\protect\citeauthoryear{{Bottrell}, {Simard}, {Mendel}  \&
  {Ellison}}{{Bottrell} et~al.}{2019}]{bottrelletal2019}
{Bottrell} C.,  {Simard} L.,  {Mendel} J.~T.,   {Ellison} S.~L.,  2019, \mn@doi
  [\mnras] {10.1093/mnras/stz855}, \href
  {https://ui.adsabs.harvard.edu/abs/2019MNRAS.486..390B} {486, 390}

\bibitem[\protect\citeauthoryear{{Bradford}, {Geha}  \& {Blanton}}{{Bradford}
  et~al.}{2015}]{lowmassgals_sdss_bradfordetal_2015}
{Bradford} J.~D.,  {Geha} M.~C.,   {Blanton} M.~R.,  2015, \mn@doi [\apj]
  {10.1088/0004-637X/809/2/146}, \href
  {https://ui.adsabs.harvard.edu/abs/2015ApJ...809..146B} {809, 146}

\bibitem[\protect\citeauthoryear{Chapon, Mayer  \& Teyssier}{Chapon
  et~al.}{2013}]{chapon_2013_maybe_gas_doesnt_always_work}
Chapon D.,  Mayer L.,   Teyssier R.,  2013, \mn@doi [Monthly Notices of the
  Royal Astronomical Society] {10.1093/mnras/sts568}, 429, 3114

\bibitem[\protect\citeauthoryear{{Cluver} et~al.,}{{Cluver}
  et~al.}{2014}]{Culver_2014}
{Cluver} M.~E.,  et~al., 2014, \mn@doi [\apj] {10.1088/0004-637X/782/2/90},
  \href {https://ui.adsabs.harvard.edu/abs/2014ApJ...782...90C} {782, 90}

\bibitem[\protect\citeauthoryear{Colpi et~al.,}{Colpi et~al.}{2019}]{colpi2019}
Colpi M.,  et~al., 2019, Astro2020 science white paper: The gravitational wave
  view of massive black holes (\mn@eprint {arXiv} {1903.06867})

\bibitem[\protect\citeauthoryear{Colpi et~al.,}{Colpi
  et~al.}{2024}]{colpi2024lisadefinitionstudyreport}
Colpi M.,  et~al., 2024, LISA Definition Study Report (\mn@eprint {arXiv}
  {2402.07571}), \url {https://arxiv.org/abs/2402.07571}

\bibitem[\protect\citeauthoryear{Comparat et~al.,}{Comparat
  et~al.}{2019}]{eboss_firefly_VAC}
Comparat J.,  et~al., 2019, Stellar population properties for 2 million
  galaxies from SDSS DR14 and DEEP2 DR4 from full spectral fitting (\mn@eprint
  {arXiv} {1711.06575}), \url {https://arxiv.org/abs/1711.06575}

\bibitem[\protect\citeauthoryear{Corrales, Haiman  \& MacFadyen}{Corrales
  et~al.}{2010}]{Corrales_2010}
Corrales L.~R.,  Haiman Z.,   MacFadyen A.,  2010, \mn@doi [Monthly Notices of
  the Royal Astronomical Society] {10.1111/j.1365-2966.2010.16324.x}, 404, 947

\bibitem[\protect\citeauthoryear{Cutler}{Cutler}{1998}]{Cutler_1998_oldlisadesign}
Cutler C.,  1998, \mn@doi [Physical Review D] {10.1103/physrevd.57.7089}, 57,
  7089–7102

\bibitem[\protect\citeauthoryear{{D'Orazio} \& {Charisi}}{{D'Orazio} \&
  {Charisi}}{2023}]{role_of_gas_in_evo}
{D'Orazio} D.~J.,  {Charisi} M.,  2023, \mn@doi [arXiv e-prints]
  {10.48550/arXiv.2310.16896}, \href
  {https://ui.adsabs.harvard.edu/abs/2023arXiv231016896D} {p. arXiv:2310.16896}

\bibitem[\protect\citeauthoryear{Dotti, Colpi, Haardt  \& Mayer}{Dotti
  et~al.}{2007}]{dotti_2007_gas_does_work}
Dotti M.,  Colpi M.,  Haardt F.,   Mayer L.,  2007, \mn@doi [Monthly Notices of
  the Royal Astronomical Society] {10.1111/j.1365-2966.2007.12010.x}, 379, 956

\bibitem[\protect\citeauthoryear{{Ellison}, {Patton}, {Mendel}  \&
  {Scudder}}{{Ellison} et~al.}{2011}]{ellison_2011}
{Ellison} S.~L.,  {Patton} D.~R.,  {Mendel} J.~T.,   {Scudder} J.~M.,  2011,
  \mn@doi [\mnras] {10.1111/j.1365-2966.2011.19624.x}, \href
  {https://ui.adsabs.harvard.edu/abs/2011MNRAS.418.2043E} {418, 2043}

\bibitem[\protect\citeauthoryear{Ellison, Mendel, Patton  \& Scudder}{Ellison
  et~al.}{2013}]{Ellison_2013}
Ellison S.~L.,  Mendel J.~T.,  Patton D.~R.,   Scudder J.~M.,  2013, \mn@doi
  [Monthly Notices of the Royal Astronomical Society] {10.1093/mnras/stt1562},
  435, 3627–3638

\bibitem[\protect\citeauthoryear{Escala, Larson, Coppi  \& Mardones}{Escala
  et~al.}{2005}]{Escala_2005_gas_does_work}
Escala A.,  Larson R.~B.,  Coppi P.~S.,   Mardones D.,  2005, \mn@doi [The
  Astrophysical Journal] {10.1086/431747}, 630, 152

\bibitem[\protect\citeauthoryear{Giacomazzo, Baker, Miller, Reynolds  \& van
  Meter}{Giacomazzo et~al.}{2012}]{Giacomazzo_2012}
Giacomazzo B.,  Baker J.~G.,  Miller M.~C.,  Reynolds C.~S.,   van Meter J.~R.,
   2012, \mn@doi [The Astrophysical Journal] {10.1088/2041-8205/752/1/l15},
  752, L15

\bibitem[\protect\citeauthoryear{{Goldstein} et~al.,}{{Goldstein}
  et~al.}{2017}]{goldstein17}
{Goldstein} A.,  et~al., 2017, \mn@doi [\apjl] {10.3847/2041-8213/aa8f41},
  \href {https://ui.adsabs.harvard.edu/abs/2017ApJ...848L..14G} {848, L14}

\bibitem[\protect\citeauthoryear{{Greene}, {Strader}  \& {Ho}}{{Greene}
  et~al.}{2020}]{greeneetal_2020_dwarfgalmassrange}
{Greene} J.~E.,  {Strader} J.,   {Ho} L.~C.,  2020, \mn@doi [\araa]
  {10.1146/annurev-astro-032620-021835}, \href
  {https://ui.adsabs.harvard.edu/abs/2020ARA&A..58..257G} {58, 257}

\bibitem[\protect\citeauthoryear{{Hickox} \& {Alexander}}{{Hickox} \&
  {Alexander}}{2018}]{hickoxetal2018}
{Hickox} R.~C.,  {Alexander} D.~M.,  2018, \mn@doi [\araa]
  {10.1146/annurev-astro-081817-051803}, \href
  {https://ui.adsabs.harvard.edu/abs/2018ARA&A..56..625H} {56, 625}

\bibitem[\protect\citeauthoryear{{Ho}}{{Ho}}{2009}]{ho_etal_2009}
{Ho} L.~C.,  2009, \mn@doi [\apj] {10.1088/0004-637X/699/1/626}, \href
  {https://ui.adsabs.harvard.edu/abs/2009ApJ...699..626H} {699, 626}

\bibitem[\protect\citeauthoryear{Hogg, Baldry, Blanton  \& Eisenstein}{Hogg
  et~al.}{2002}]{hogg2002k}
Hogg D.~W.,  Baldry I.~K.,  Blanton M.~R.,   Eisenstein D.~J.,  2002, The K
  correction (\mn@eprint {arXiv} {astro-ph/0210394})

\bibitem[\protect\citeauthoryear{Holley-Bockelmann \& Khan}{Holley-Bockelmann
  \& Khan}{2015}]{Holley_Bockelmann_2015_bhs_merge_timescale_long}
Holley-Bockelmann K.,  Khan F.~M.,  2015, \mn@doi [The Astrophysical Journal]
  {10.1088/0004-637x/810/2/139}, 810, 139

\bibitem[\protect\citeauthoryear{Holz \& Hughes}{Holz \&
  Hughes}{2005}]{Holz_2005}
Holz D.~E.,  Hughes S.~A.,  2005, \mn@doi [The Astrophysical Journal]
  {10.1086/431341}, 629, 15

\bibitem[\protect\citeauthoryear{{Hughes}}{{Hughes}}{2002}]{hughes2002}
{Hughes} S.~A.,  2002, \mn@doi [\mnras] {10.1046/j.1365-8711.2002.05247.x},
  \href {https://ui.adsabs.harvard.edu/abs/2002MNRAS.331..805H} {331, 805}

\bibitem[\protect\citeauthoryear{{Ivezi{\'c}} et~al.,}{{Ivezi{\'c}}
  et~al.}{2019}]{Ivezic_2019_LSST}
{Ivezi{\'c}} {\v{Z}}.,  et~al., 2019, \mn@doi [\apj]
  {10.3847/1538-4357/ab042c}, \href
  {https://ui.adsabs.harvard.edu/abs/2019ApJ...873..111I} {873, 111}

\bibitem[\protect\citeauthoryear{Izquierdo-Villalba, Colpi, Volonteri, Spinoso,
  Bonoli  \& Sesana}{Izquierdo-Villalba et~al.}{2023}]{izetal2023}
Izquierdo-Villalba D.,  Colpi M.,  Volonteri M.,  Spinoso D.,  Bonoli S.,
  Sesana A.,  2023, \mn@doi [A\&A] {10.1051/0004-6361/202347008}, 677, A123

\bibitem[\protect\citeauthoryear{{Jarrett} et~al.,}{{Jarrett}
  et~al.}{2013}]{Jarrett_2013}
{Jarrett} T.~H.,  et~al., 2013, \mn@doi [\aj] {10.1088/0004-6256/145/1/6},
  \href {https://ui.adsabs.harvard.edu/abs/2013AJ....145....6J} {145, 6}

\bibitem[\protect\citeauthoryear{Jarrett, Cluver, Taylor, Bellstedt, Robotham
  \& Yao}{Jarrett et~al.}{2023}]{lowmassgals_WISE_jarrettetal_2023}
Jarrett T.~H.,  Cluver M.~E.,  Taylor E.~N.,  Bellstedt S.,  Robotham A. S.~G.,
    Yao H. F.~M.,  2023, A New WISE Calibration of Stellar Mass (\mn@eprint
  {arXiv} {2301.05952}), \url {https://arxiv.org/abs/2301.05952}

\bibitem[\protect\citeauthoryear{Jiang et~al.,}{Jiang
  et~al.}{2014}]{Jiang_2014_s82}
Jiang L.,  et~al., 2014, \mn@doi [The Astrophysical Journal Supplement Series]
  {10.1088/0067-0049/213/1/12}, 213, 12

\bibitem[\protect\citeauthoryear{Katz, Marsat, Chua, Babak  \& Larson}{Katz
  et~al.}{2020}]{Katz_2020}
Katz M.~L.,  Marsat S.,  Chua A.~J.,  Babak S.,   Larson S.~L.,  2020, \mn@doi
  [Physical Review D] {10.1103/physrevd.102.023033}, 102

\bibitem[\protect\citeauthoryear{Kelly, Baker, Etienne, Giacomazzo  \&
  Schnittman}{Kelly et~al.}{2017}]{Kelly_2017}
Kelly B.~J.,  Baker J.~G.,  Etienne Z.~B.,  Giacomazzo B.,   Schnittman J.,
  2017, \mn@doi [Physical Review D] {10.1103/physrevd.96.123003}, 96

\bibitem[\protect\citeauthoryear{{Kettlety} et~al.,}{{Kettlety}
  et~al.}{2018}]{kettlety18}
{Kettlety} T.,  et~al., 2018, \mn@doi [\mnras] {10.1093/mnras/stx2379}, \href
  {https://ui.adsabs.harvard.edu/abs/2018MNRAS.473..776K} {473, 776}

\bibitem[\protect\citeauthoryear{Khan, Capelo, Mayer  \& Berczik}{Khan
  et~al.}{2018}]{Khan_2018_bhswillmerge}
Khan F.~M.,  Capelo P.~R.,  Mayer L.,   Berczik P.,  2018, \mn@doi [The
  Astrophysical Journal] {10.3847/1538-4357/aae77b}, 868, 97

\bibitem[\protect\citeauthoryear{Kimbrell, Reines, Greene  \& Geha}{Kimbrell
  et~al.}{2023}]{Kimbrell_2023_bd_decon_in_dwarf}
Kimbrell S.~J.,  Reines A.~E.,  Greene J.~E.,   Geha M.,  2023, \mn@doi [The
  Astrophysical Journal] {10.3847/1538-4357/acf762}, 958, 115

\bibitem[\protect\citeauthoryear{Klein et~al.,}{Klein
  et~al.}{2016}]{kleinetal2016}
Klein A.,  et~al., 2016, \mn@doi [Phys. Rev. D] {10.1103/PhysRevD.93.024003},
  93, 024003

\bibitem[\protect\citeauthoryear{Kocsis, Frei, Haiman  \& Menou}{Kocsis
  et~al.}{2006}]{Kocsis_2006}
Kocsis B.,  Frei Z.,  Haiman Z.,   Menou K.,  2006, \mn@doi [The Astrophysical
  Journal] {10.1086/498236}, 637, 27–37

\bibitem[\protect\citeauthoryear{Kocsis, Haiman  \& Menou}{Kocsis
  et~al.}{2008}]{Kocsis_2008}
Kocsis B.,  Haiman Z.,   Menou K.,  2008, \mn@doi [The Astrophysical Journal]
  {10.1086/590230}, 684, 870

\bibitem[\protect\citeauthoryear{{Lang} \& {Hughes}}{{Lang} \&
  {Hughes}}{2008}]{langetal2008}
{Lang} R.~N.,  {Hughes} S.~A.,  2008, \mn@doi [\apj] {10.1086/528953}, \href
  {https://ui.adsabs.harvard.edu/abs/2008ApJ...677.1184L} {677, 1184}

\bibitem[\protect\citeauthoryear{{Leroy} et~al.,}{{Leroy}
  et~al.}{2019}]{leroyetal2019}
{Leroy} A.~K.,  et~al., 2019, \mn@doi [\apjs] {10.3847/1538-4365/ab3925}, \href
  {https://ui.adsabs.harvard.edu/abs/2019ApJS..244...24L} {244, 24}

\bibitem[\protect\citeauthoryear{Lima, Mayer, Capelo, Bortolas  \& Quinn}{Lima
  et~al.}{2020}]{limaetal_2020}
Lima R.~S.,  Mayer L.,  Capelo P.~R.,  Bortolas E.,   Quinn T.~R.,  2020, The
  erratic path to coalescence of LISA massive black hole binaries in sub-pc
  resolution simulations of smooth circumnuclear gas disks (\mn@eprint {arXiv}
  {2003.13789}), \url {https://arxiv.org/abs/2003.13789}

\bibitem[\protect\citeauthoryear{Lippai, Frei  \& Haiman}{Lippai
  et~al.}{2008}]{Lippai_2008}
Lippai Z.,  Frei Z.,   Haiman Z.,  2008, \mn@doi [The Astrophysical Journal]
  {10.1086/587034}, 676, L5

\bibitem[\protect\citeauthoryear{Lodato, Nayakshin, King  \& Pringle}{Lodato
  et~al.}{2009}]{lodato_2009_gas_does_work_intricate}
Lodato G.,  Nayakshin S.,  King A.~R.,   Pringle J.~E.,  2009, \mn@doi [Monthly
  Notices of the Royal Astronomical Society]
  {10.1111/j.1365-2966.2009.15179.x}, 398, 1392

\bibitem[\protect\citeauthoryear{Lops, Izquierdo-Villalba, Colpi, Bonoli,
  Sesana  \& Mangiagli}{Lops et~al.}{2023}]{lopsetal2023}
Lops G.,  Izquierdo-Villalba D.,  Colpi M.,  Bonoli S.,  Sesana A.,   Mangiagli
  A.,  2023, \mn@doi [Monthly Notices of the Royal Astronomical Society]
  {10.1093/mnras/stad058}, 519, 5962

\bibitem[\protect\citeauthoryear{{Mangiagli} et~al.,}{{Mangiagli}
  et~al.}{2020}]{mangiagli20}
{Mangiagli} A.,  et~al., 2020, \mn@doi [\prd] {10.1103/PhysRevD.102.084056},
  \href {https://ui.adsabs.harvard.edu/abs/2020PhRvD.102h4056M} {102, 084056}

\bibitem[\protect\citeauthoryear{{Mangiagli}, {Caprini}, {Marsat}, {Speri},
  {Caldwell}  \& {Tamanini}}{{Mangiagli} et~al.}{2025}]{mangiagli2025}
{Mangiagli} A.,  {Caprini} C.,  {Marsat} S.,  {Speri} L.,  {Caldwell} R.~R.,
  {Tamanini} N.,  2025, \mn@doi [\prd] {10.1103/PhysRevD.111.083043}, \href
  {https://ui.adsabs.harvard.edu/abs/2025PhRvD.111h3043M} {111, 083043}

\bibitem[\protect\citeauthoryear{McConnell \& Ma}{McConnell \&
  Ma}{2013}]{mcconnell2013}
McConnell N.~J.,  Ma C.-P.,  2013, \mn@doi [The Astrophysical Journal]
  {10.1088/0004-637x/764/2/184}, 764, 184

\bibitem[\protect\citeauthoryear{{Medling} et~al.,}{{Medling}
  et~al.}{2015}]{Medling_2015_return_to_relation_timescale}
{Medling} A.~M.,  et~al., 2015, \mn@doi [\apj] {10.1088/0004-637X/803/2/61},
  \href {https://ui.adsabs.harvard.edu/abs/2015ApJ...803...61M} {803, 61}

\bibitem[\protect\citeauthoryear{Meliani, Mizuno, Olivares, Porth, Rezzolla  \&
  Younsi}{Meliani et~al.}{2017}]{Meliani_2017}
Meliani Z.,  Mizuno Y.,  Olivares H.,  Porth O.,  Rezzolla L.,   Younsi Z.,
  2017, \mn@doi [Astronomy \& Astrophysics] {10.1051/0004-6361/201629191}, 598,
  A38

\bibitem[\protect\citeauthoryear{{Mendel}, {Simard}, {Palmer}, {Ellison}  \&
  {Patton}}{{Mendel} et~al.}{2014}]{mendeletal2014}
{Mendel} J.~T.,  {Simard} L.,  {Palmer} M.,  {Ellison} S.~L.,   {Patton} D.~R.,
   2014, \mn@doi [\apjs] {10.1088/0067-0049/210/1/3}, \href
  {https://ui.adsabs.harvard.edu/abs/2014ApJS..210....3M} {210, 3}

\bibitem[\protect\citeauthoryear{Merloni et~al.,}{Merloni
  et~al.}{2013}]{merlonietal2013}
Merloni A.,  et~al., 2013, \mn@doi [Monthly Notices of the Royal Astronomical
  Society] {10.1093/mnras/stt2149}, 437, 3550

\bibitem[\protect\citeauthoryear{Moesta, Alic, Rezzolla, Zanotti  \&
  Palenzuela}{Moesta et~al.}{2012}]{Moesta_2012}
Moesta P.,  Alic D.,  Rezzolla L.,  Zanotti O.,   Palenzuela C.,  2012, \mn@doi
  [The Astrophysical Journal] {10.1088/2041-8205/749/2/l32}, 749, L32

\bibitem[\protect\citeauthoryear{Molina, Reines, Latimer, Baldassare  \&
  Salehirad}{Molina et~al.}{2021}]{Molina_2021_agnindwarfs}
Molina M.,  Reines A.~E.,  Latimer L.,  Baldassare V.,   Salehirad S.,  2021,
  \mn@doi [The Astrophysical Journal] {10.3847/1538-4357/ac1ffa}, 922, 155

\bibitem[\protect\citeauthoryear{Palenzuela, Lehner  \& Liebling}{Palenzuela
  et~al.}{2010}]{Palenzuela_2010}
Palenzuela C.,  Lehner L.,   Liebling S.~L.,  2010, \mn@doi [Science]
  {10.1126/science.1191766}, 329, 927

\bibitem[\protect\citeauthoryear{{Peng}, {Ho}, {Impey}  \& {Rix}}{{Peng}
  et~al.}{2010}]{peng_2010_gal_morpho_bd}
{Peng} C.~Y.,  {Ho} L.~C.,  {Impey} C.~D.,   {Rix} H.-W.,  2010, \mn@doi [\aj]
  {10.1088/0004-6256/139/6/2097}, \href
  {https://ui.adsabs.harvard.edu/abs/2010AJ....139.2097P} {139, 2097}

\bibitem[\protect\citeauthoryear{Reines}{Reines}{2022}]{Reines_2022_dwarfgalmassrange}
Reines A.~E.,  2022, \mn@doi [Nature Astronomy] {10.1038/s41550-021-01556-0},
  6, 26–34

\bibitem[\protect\citeauthoryear{{Reines}, {Greene}  \& {Geha}}{{Reines}
  et~al.}{2013}]{reinesetal_2013_agnindwarfs}
{Reines} A.~E.,  {Greene} J.~E.,   {Geha} M.,  2013, \mn@doi [\apj]
  {10.1088/0004-637X/775/2/116}, \href
  {https://ui.adsabs.harvard.edu/abs/2013ApJ...775..116R} {775, 116}

\bibitem[\protect\citeauthoryear{Rossi, Lodato, Armitage, Pringle  \&
  King}{Rossi et~al.}{2010}]{Rossi_2010}
Rossi E.~M.,  Lodato G.,  Armitage P.~J.,  Pringle J.~E.,   King A.~R.,  2010,
  \mn@doi [Monthly Notices of the Royal Astronomical Society]
  {10.1111/j.1365-2966.2009.15802.x}, 401, 2021

\bibitem[\protect\citeauthoryear{{Savchenko} et~al.,}{{Savchenko}
  et~al.}{2017}]{savchenko17}
{Savchenko} V.,  et~al., 2017, \mn@doi [\apjl] {10.3847/2041-8213/aa8f94},
  \href {https://ui.adsabs.harvard.edu/abs/2017ApJ...848L..15S} {848, L15}

\bibitem[\protect\citeauthoryear{Schnittman \& Krolik}{Schnittman \&
  Krolik}{2008}]{Schnittman_2008}
Schnittman J.~D.,  Krolik J.~H.,  2008, \mn@doi [The Astrophysical Journal]
  {10.1086/590363}, 684, 835

\bibitem[\protect\citeauthoryear{{Schutte}, {Reines}  \& {Greene}}{{Schutte}
  et~al.}{2019}]{schutte19}
{Schutte} Z.,  {Reines} A.~E.,   {Greene} J.~E.,  2019, \mn@doi [\apj]
  {10.3847/1538-4357/ab35dd}, \href
  {https://ui.adsabs.harvard.edu/abs/2019ApJ...887..245S} {887, 245}

\bibitem[\protect\citeauthoryear{{Simard}, {Mendel}, {Patton}, {Ellison}  \&
  {McConnachie}}{{Simard} et~al.}{2011}]{simardetal2011}
{Simard} L.,  {Mendel} J.~T.,  {Patton} D.~R.,  {Ellison} S.~L.,
  {McConnachie} A.~W.,  2011, \mn@doi [\apjs] {10.1088/0067-0049/196/1/11},
  \href {https://ui.adsabs.harvard.edu/abs/2011ApJS..196...11S} {196, 11}

\bibitem[\protect\citeauthoryear{{Sun}, {Barger}, {Frinchaboy}  \& {Pan}}{{Sun}
  et~al.}{2020}]{lowmassglas_sdss_sunetal_2020}
{Sun} J.,  {Barger} K.~A.,  {Frinchaboy} P.~M.,   {Pan} K.,  2020, \mn@doi
  [\apj] {10.3847/1538-4357/ab6dee}, \href
  {https://ui.adsabs.harvard.edu/abs/2020ApJ...894...57S} {894, 57}

\bibitem[\protect\citeauthoryear{Tamanini}{Tamanini}{2017}]{Tamanini_2017}
Tamanini N.,  2017, \mn@doi [Journal of Physics: Conference Series]
  {10.1088/1742-6596/840/1/012029}, 840, 012029

\bibitem[\protect\citeauthoryear{Vecchio}{Vecchio}{2004}]{Vecchio_2004_oldparamest}
Vecchio A.,  2004, \mn@doi [Physical Review D] {10.1103/physrevd.70.042001}, 70

\bibitem[\protect\citeauthoryear{{Wilkinson}, {Maraston}, {Goddard}, {Thomas}
  \& {Parikh}}{{Wilkinson} et~al.}{2017}]{wilkinson_2019_firefly_code}
{Wilkinson} D.~M.,  {Maraston} C.,  {Goddard} D.,  {Thomas} D.,   {Parikh} T.,
  2017, \mn@doi [\mnras] {10.1093/mnras/stx2215}, \href
  {https://ui.adsabs.harvard.edu/abs/2017MNRAS.472.4297W} {472, 4297}

\bibitem[\protect\citeauthoryear{{Wright} et~al.,}{{Wright}
  et~al.}{2010}]{WISE}
{Wright} E.~L.,  et~al., 2010, \mn@doi [\aj] {10.1088/0004-6256/140/6/1868},
  \href {https://ui.adsabs.harvard.edu/abs/2010AJ....140.1868W} {140, 1868}

\bibitem[\protect\citeauthoryear{{York} et~al.,}{{York} et~al.}{2000}]{SDSS}
{York} D.~G.,  et~al., 2000, \mn@doi [\aj] {10.1086/301513}, \href
  {https://ui.adsabs.harvard.edu/abs/2000AJ....120.1579Y} {120, 1579}

\bibitem[\protect\citeauthoryear{Yu et~al.,}{Yu et~al.}{2025}]{Yu_2025}
Yu W.,  et~al., 2025, \mn@doi [The Astrophysical Journal]
  {10.3847/1538-4357/adb283}, 981, 141

\makeatother
\end{thebibliography}




\bsp	
\label{lastpage}
\end{document}